\documentclass[%
 reprint,
 amsmath,amssymb,
 aps,
]{revtex4-2}

\usepackage{graphicx}% Include figure files
\usepackage{dcolumn}% Align table columns on decimal point
\usepackage{bm}% bold math
\usepackage{color}
\usepackage{verbatim}
\usepackage{longtable}
\usepackage{multirow}
\usepackage{appendix}

\begin{document}

\preprint{APS/123-QED}

\title{Measurement of the Low-Frequency Charge Noise of Bacteria} % Force line breaks with \\
%\thanks{A footnote to the article title}%

\author{Yichao Yang}
% \altaffiliation[Also at ]{Physics Department, XYZ University.}%Lines break automatically or can be forced with \\
\author{Hagen Gress}
% \altaffiliation[Also at ]{Physics Department, XYZ University.}%Lines break automatically or can be forced with \\
\author{Kamil L. Ekinci}%
\email[Electronic mail: ]{ekinci@bu.edu}
% \email{Second.Author@institution.edu}
\affiliation{Department of Mechanical Engineering, Division of Materials Science and Engineering, and the Photonics Center, Boston University, Boston, Massachusetts 02215, United States}

\date{\today}% It is always \today, today,
             %  but any date may be explicitly specified

\begin{abstract}

Bacteria meticulously regulate their intracellular ion concentrations and create ionic concentration gradients across the bacterial membrane. These ionic concentration gradients  provide free energy for many cellular processes and are maintained by transmembrane transport.   Given the physical dimensions of a bacterium and the stochasticity  in transmembrane transport, intracellular  ion  concentrations  and hence the charge state of a bacterium   are bound to fluctuate.  Here, we investigate the   charge noise  of 100s of non-motile bacteria  by combining electrical  measurement techniques from condensed matter physics  with microfluidics. In our experiments,   bacteria   in a microchannel  generate  charge  density fluctuations in the embedding electrolyte  due to random influx and efflux of ions. Detected  as electrical resistance noise, these charge density fluctuations  display a power spectral density proportional to $1/f^2$ for frequencies $0.05~{\rm Hz} \leq f \leq 1 ~{\rm Hz}$. Fits  to a simple  noise model suggest that the steady-state  charge  of a  bacterium fluctuates by  $\pm 1.30 \times 10^6 {e}~({e} \approx 1.60 \times 10^{-19}~{\rm C})$, indicating that  bacterial ion homeostasis is highly dynamic and  dominated by strong charge noise.  The rms charge noise can then be used to estimate the fluctuations in the membrane potential; however, the  estimates are unreliable due to our limited understanding of the intracellular concentration gradients.

\end{abstract}

%\keywords{Suggested keywords}%Use showkeys class option if keyword
                              %display desired
\maketitle

%\tableofcontents

\section{Introduction}

\begin{figure*}
\centering
\includegraphics[width=6.75in]{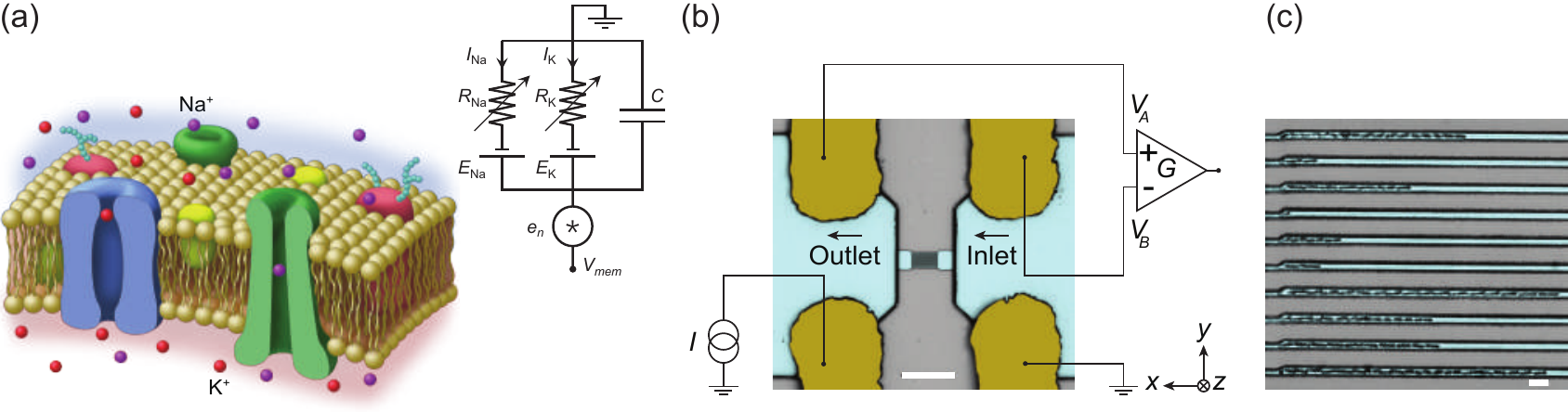}% Here is how to import EPS art
\caption{ (a) Illustration of a membrane patch and its circuit model. (b) Microfluidic resistor and the simplified circuit diagram  for monitoring its electrical fluctuations. The microchannels in the center of the resistor are filled with non-motile bacteria. The false-colored inverted microscope image  shows the electrodes (gold) and the broth medium (light blue) looking from the bottom glass side of the device. The scale bar is 200 $\mu$m. Arrows indicate the direction of the flow. The current source pushes a bias current through the device. The ensuing voltage drop, $V_A-V_B$, is detected by a differential amplifier  with gain $G$. (See Fig. \ref{fig:figure2}(a) below for more details.)  (c) Close up image of the microchannels filled with \textit{K. pneumoniae} cells. Each microchannel has linear dimensions of $l\times w\times h\approx 100\times 2\times 2 ~\mu \rm m^3$; the cross-section reduces to  $800~\rm nm \times 2 ~\mu \rm m$ at the constriction. The scale bar is 5 $\mu$m. }
\label{fig:figure1}
\end{figure*}

Bacteria  create and maintain   transmembrane  concentration gradients of small metal ions, such as potassium and sodium \cite{galera2021ionobiology, korolev2021potassium}.  These ionic concentration gradients are the main source of the electrical and electrochemical potentials that are present across the cell membrane and  facilitate a number of crucial cellular processes  \cite{bruni2017voltage, anishkin2014feeling, prindle2015ion, mitchell1961coupling,strahl2010membrane, dong2019magnesium, damper1981role}. The processes for  separation, concentration and regulation of the inorganic ions  by bacteria  are made possible by the plasma membrane and  the transmembrane proteins embedded in the membrane, as shown in Fig. \ref{fig:figure1}(a).    The plasma  membrane is a thin but strongly insulating lipid structure, allowing a bacterium  to maintain  its charge state effectively. The  many different  transmembrane proteins that are  in the bacterial membrane act as channels and pumps for ions  \cite{gadsby2009ion}.  Figure \ref{fig:figure1}(a) also shows  the electrical circuit model \cite{prindle2015ion, dong2019magnesium, yang2020encoding} of a  membrane patch. Here, the ion  channels and pumps for each ion are modeled as a nonlinear  resistor with a conductance that depends on the electrical potential of the membrane, $V_{mem}$, and the Nernst potential, $E_{\rm X}$, for the ion $\rm X$.   

A small and insulating system, such as a bacterium, will be particularly susceptible to charge fluctuations \cite{benarroch2020microbiologist, schofield2020bioelectrical,kralj2011electrical}. Thus, charge noise should play an important role in bacterial ion homeostasis. Based on the circuit model [Fig. 1(a)],   the  charge fluctuations within the cytoplasm are the result of  two coupled noisy processes. First, the ionic current through the ion channels is noisy  \cite{queralt2020specific}, with the ensuing fluctuations   in intracellular ion concentrations  causing voltage noise in $V_{mem}$. Second, any noise originating in $V_{mem}$, e.g., due to a  random depolarization of a membrane patch or even thermal noise \cite{defelice1981noise,weaver1990response}, will  cause fluctuations in the transmembrane current \cite{xie1997fluctuation} and hence the intracellular ion concentrations. Furthermore, these two noise processes  will tend to   enhance  each other.  We argue that an equivalent noise voltage, $e_n$, should be imposed on $V_{mem}$ in Fig. \ref{fig:figure1}(a)  in order to account for all the electrical fluctuations  present. 

Our understanding of how a bacterium regulates the concentrations  of metal ions in its cytoplasm (i.e., the intracellular metallome) is far from complete. It is generally assumed that the time-averaged intracellular ion concentrations in a bacterium remain  roughly constant \cite{endresen2000theory, hund2001ionic}---indicating charge conservation.   Average bacterial ion efflux \cite{shabala2001measurements} and influx \cite{shabala2009ion} rates have  been determined from  concentration measurements in  media in which large populations of bacteria are grown. On the other hand, patch clamp measurements on single bacterial ion channels  have shown that  ion transport is noisy, with the current noise power spectral density (PSD)  proportional to  $1/f$ at low frequencies \cite{bezrukov2000examining}.   While the time-averaged    charge state remains constant,  bacteria surprisingly modulate their membrane potential on the timescale of seconds---as shown in recent fluorescent microscopy experiments \cite{kralj2011electrical}. Single  cells    get hyperpolarized and depolarized spontaneously and repeatedly over time. It has been suggested that these charge fluctuations are purposeful, but little is known about their downstream effects. Thus, bacterial ion homeostasis is expected to be highly dynamic \cite{galera2021ionobiology} and  dominated by strong charge noise. At the present time, quantitative measurements of electrical fluctuations and   noise models of single bacterial cells are missing from the biophysics literature,  partly due to a lack of sensitive tools   on the size scale of a bacterium \cite{kralj2011electrical}.  

%*************************************Figure************************
%%%%%%%%%%%%%%%%%%%%%%%%%%%%%%%%%%%%%%%%%%%%%%%%%%%%%%%%%%%%%%%%%
%%%%%%%%%%%%%%%%%%%%%%%%%%%%%%%%%%%%%%%%%%%%%%%%%%%%%%%%%%%%%%%%%
%%%%%%%%%%%%%%%%%%%%%%%%%%%%%%%%%%%%%%%%%%%%%%%%%%%%%%%%%%%%%%%%%
%%%%%%%%%%%%%%%%%%%%%%%%%%%%%%%%%%%%%%%%%%%%%%%%%%%%%%%%%%%%%%%%%

To probe  the noisy charge dynamics in bacteria with sufficient time resolution and electrical sensitivity,  we have combined low-frequency noise analyses from  condensed matter physics \cite{scofield1987ac} with microfluidics \cite{yang2020all}. Our  overarching hypothesis is that the metabolic activity of bacteria
modulates the  concentrations of various ions in the medium, leading to detectable fluctuations in the electrical impedance. In Section \ref{sec:approach}, we present our experimental approach with particular attention to the electrical measurements. The results in Section \ref{sec:results} establish that  electrical fluctuations detected from bacteria  scale as  $1/f^2$, with the characteristics of equilibrium resistance noise \cite{voss1976flicker, dutta1981low, verleg1998resistance, tasserit2010pink, wen2017generalized}. In Section \ref{sec:discussion}, we look at possible noise mechanisms and discuss the possibility of charge noise. Section \ref{sec:conclusion} is reserved for conclusions. In Appendix \ref{app:experimental}, we present further experimental details. In Appendix \ref{app:dimensionless}, we discuss how to consistently remove the superficial effects of the measurement circuit from the  noise.  In Appendix \ref{app:perturbations}, we present results from  our control experiments and discuss perturbations; we also discuss the contribution of another phenomenon, namely the nanomechanical fluctuations of a bacterium, to the observed  noise. In Appendix \ref{app:chargenoise}, we present the details of how we estimate various electrical noise quantities.

\section{Experimental Approach}
\label{sec:approach}

\subsection{Device and setup}

\subsubsection{Microfluidic resistor}

We perform our electrical noise measurements in a PDMS microfluidic resistor that sits on an inverted microscope stage.    As shown in Fig. \ref{fig:figure1}(b), the device  has ten parallel microchannels at its center, each with a  nanoscale constriction toward the outlet  end [Fig. \ref{fig:figure1}(c)]. The  microfluidic resistor is filled with a liquid electrolyte, such as Luria-Bertani (LB) broth medium or phosphate-buffered saline (PBS),  and four thin film Cr-Au  electrodes  allow for electrical contact to the ions in the  medium [Fig. \ref{fig:figure1}(b)]. Cr-Au electrodes are a good alternative to AgCl electrodes: they are easy to fabricate and provide satisfactory electrical properties  \cite{rocha2016electrochemical,tsutsui2011single,maleki2009nanofluidic}. Unless otherwise noted, the temperature of the media is kept at $37^{\circ}$C by the temperature-controlled inverted microscope stage. 

%*************************************Figure************************
%%%%%%%%%%%%%%%%%%%%%%%%%%%%%%%%%%%%%%%%%%%%%%%%%%%%%%%%%%%%%%%%%
%%%%%%%%%%%%%%%%%%%%%%%%%%%%%%%%%%%%%%%%%%%%%%%%%%%%%%%%%%%%%%%%%
\begin{figure*}
    \begin{center}
       \includegraphics[width=6.75 in]{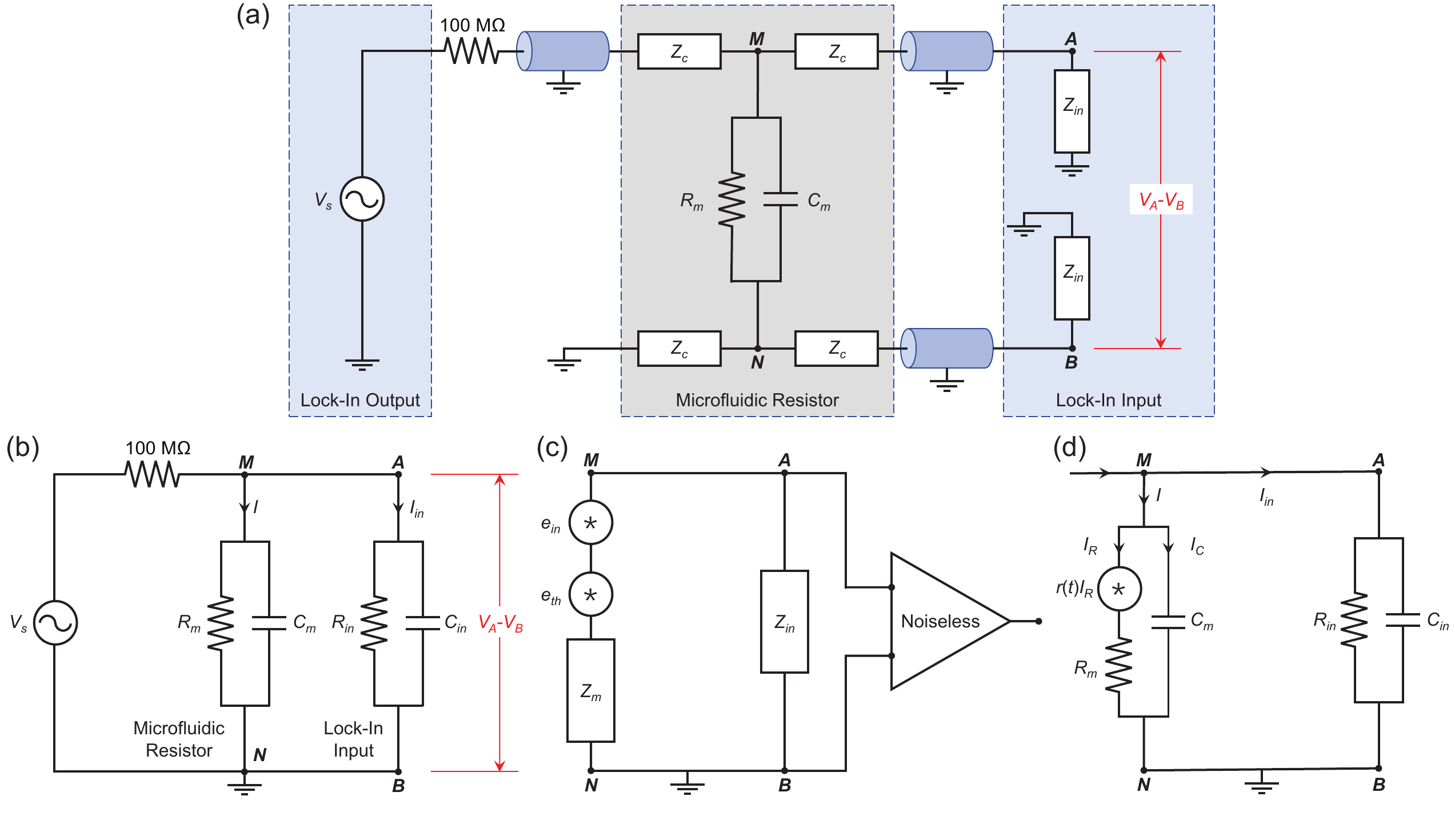}
      \caption{(a) A schematic diagram of the  circuit for  electrical measurements. The dashed boxes show the lock-in amplifier reference oscillator output (left), the microfluidic resistor (center), and the lock-in amplifier input (right). (b) A simplified equivalent circuit for the measurement, showing  the impedances of the microfludic resistor and the lock-in input. The arrows show the current flow directions. (c)  Th{\'e}venin equivalent noise circuit showing the thermal noise of the microfluidic resistor and the input noise of the lock-in amplifier. (d) Equivalent circuit showing the excess noise in the microfluidic resistor. }
    \label{fig:figure2}
    \end{center}
\end{figure*}
%%%%%%%%%%%%%%%%%%%%%%%%%%%%%%%%%%%%%%%%%%%%%%%%%%%%%%%%%%%%%%%%%
%%%%%%%%%%%%%%%%%%%%%%%%%%%%%%%%%%%%%%%%%%%%%%%%%%%%%%%%%%%%%%%%%%% 

\subsubsection{Loading and trapping the bacteria}

The noise measurements are performed  on both live and dead bacteria trapped in these  parallel microchannels.  At the start of  each experiment,  We load and trap the bacteria using a pressure-driven flow of a bacteria solution from the inlet toward the nanoconstriction with $\Delta p \approx 10~\rm kPa$. The bacteria accumulate in the microchannels in a linear fashion, as shown in Fig. 1(c). Once the bacteria are loaded, we wait for 30 min before starting the electrical measurements. During the measurements, we maintain a constant $\Delta p\approx 0.6~\rm kPa$ to ensure the flow of nutrients to the bacteria in the microchannel region, except in experiments for studying the effects of different $\Delta p$ values. The pressure-driven flow jams the bacteria towards the nanoconstriction and  keeps the bacteria from moving; it also prevents the bacteria from oscillating due to the electrokinetic forces.

\subsubsection{ Properties of the bacteria}

We use non-motile Gram-negative bacteria (\textit{Klebsiella pneumoniae}) and  Gram-positive bacteria (\textit{Staphylococcus saprophyticus}) in our experiments.  \textit{K. pneumoniae} is a rod-shaped  microorganism that has a length of $2-3 ~\mu \rm m$ and a cross-sectional area of $ 0.8 ~\mu \rm m^2$ \cite{zhang2021measurement}; \textit{S. saprophyticus} is a spherical microorganism that has a  diameter of $ 1 ~\mu \rm m$ \cite{monteiro2015cell}. The average doubling times for \textit{K. pneumoniae} and \textit{S. saprophyticus} in our microfluidic devices  at $37^{\circ}$C are  55 min. and 100 min., respectively \cite{yang2020all}.  In  experiments with dead bacteria, the cells are first killed by adding a small amount of  glutaraldehyde into the broth medium, before they are pushed into the microchannels. Glutaraldehyde is a fixative which kills bacteria by impeding essential cellular functions but  preserves the cellular morphology and ultrastructure for the  period of our experiments ($\sim3$ hours)  \cite{mcdonnell1999antiseptics}.

%*************************************Figure************************
%%%%%%%%%%%%%%%%%%%%%%%%%%%%%%%%%%%%%%%%%%%%%%%%%%%%%%%%%%%%%%%%%
%%%%%%%%%%%%%%%%%%%%%%%%%%%%%%%%%%%%%%%%%%%%%%%%%%%%%%%%%%%%%%%%%
\begin{figure*}[t!]
    \begin{center}
    \includegraphics[width=6.75 in]{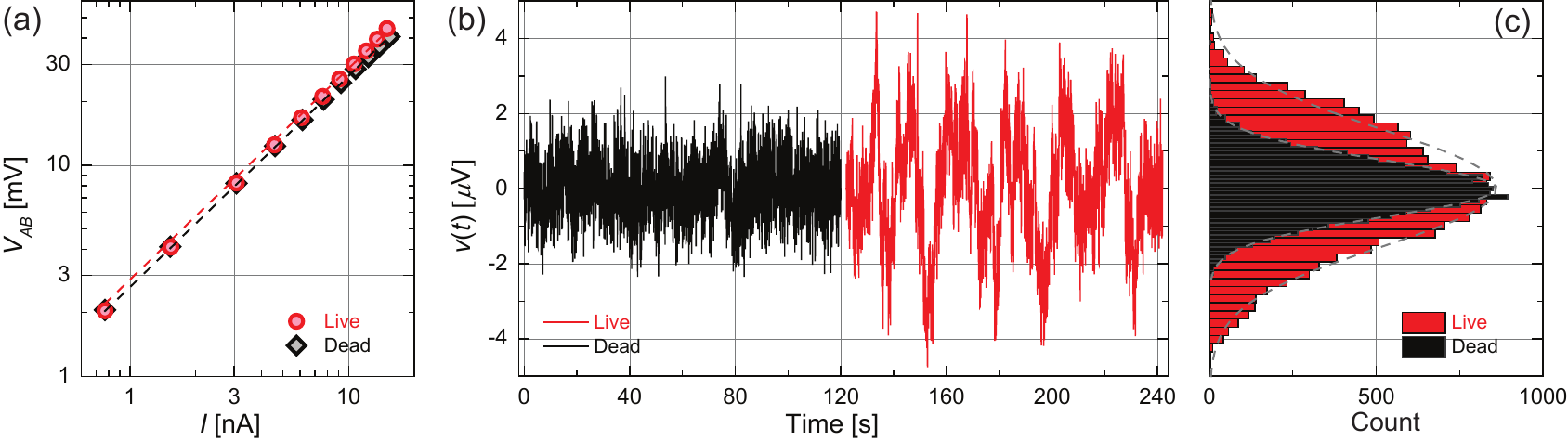} 
    \caption{(a) The  $I$-$V$ characteristics of devices filled with live  and  fixed (dead)  \textit{K. pneumoniae} cells.  The average value of the phase angle is $-34 \pm 3^{\circ}$. (b) Voltage fluctuations $v(t)$ from live and dead bacteria at a fixed $I\approx 7.74$ nA bias in a (noise) bandwidth of $\Delta f \approx 10 \rm~Hz$ as a function of time. There are $N\approx 300$ bacteria trapped in both devices. (c) Histograms. The dashed lines show Gaussians.}
    \label{fig:figure3}
    \end{center}
\end{figure*}
%%%%%%%%%%%%%%%%%%%%%%%%%%%%%%%%%%%%%%%%%%%%%%%%%%%%%%%%%%%%%%%%%
%%%%%%%%%%%%%%%%%%%%%%%%%%%%%%%%%%%%%%%%%%%%%%%%%%%%%%%%%%%%%%%%%%%

\subsection{Electrical measurements} 
\label{sec:electricalmeasurments}

\subsubsection{Lock-in amplifier setup}
We employ  a four-probe ac measurement using a lock-in amplifier, as shown schematically  in Fig. \ref{fig:figure1}(b)  and Fig. \ref{fig:figure2}(a).   In Fig. \ref{fig:figure2}(a), $V_s$ is the reference oscillator output of the lock-in amplifier, with the reference frequency  set to $f_o=160~\rm Hz$; $Z_c$ is the contact impedance at each of the four  probes; the impedance $Z_m$ of the microfluidic resistor is modeled as a resistor $R_m$ in parallel with a capacitance $C_m$; and $Z_{in}$ is the equivalent impedance of the lock-in amplifier input, with ${V_{A}}-{V_{B}}$ representing the  voltage drop between the two differential inputs $A$ and $B$. A  current source is created by connecting  the lock-in oscillator output  in series with a $100 ~\rm M\Omega$ metal film resistor that has a parasitic capacitance $\le 0.4$ pF.   The voltage drop across the microfluidic resistor is measured by the differential voltage detection mode of the lock-in amplifier [Fig. \ref{fig:figure1}(b)]. 

During the noise measurements, an ac bias current of rms amplitude in the range $ 0.70~{\rm nA} \le {I} \le 16.50~{\rm nA}$ is pushed through the microfluidic resistor, and the voltage fluctuations across the resistor are detected. The time constant and filter roll-off of the lock-in amplifier are $3~\rm ms$ and $18~\rm dB/oct$, respectively, resulting in an equivalent noise bandwidth of $31.25~\rm Hz$. The data are transferred to a computer using a digitizer at a sampling rate of $128~\rm Hz$.  Our subsequent numerical filtering  only keeps the noise in the frequency interval $0.05~{\rm Hz} \le f \le 10~\rm Hz$. To summarize, our measurements are similar to the ac noise measurements  on solid state systems but without the  bridge configuration \cite{scofield19851}. One big advantage in our system is that we can establish a background noise level by measurements on fixed (dead) bacteria that  lack any metabolic activity.

\subsubsection{Mean voltage drop and estimation of the circuit parameters}

Figure \ref{fig:figure3}(a)  shows the rms value of the  mean voltage drop, $V_{AB}={V_{A}}-{V_{B}}$, for two devices as a function of the rms bias current $I$ that flows through the devices; the  value of the phase angle is $-34 \pm 3^{\circ}$ and stays constant. The devices are nominally identical, but one is filled with live cells and the other with dead cells. To find  $I$, $R_m$ and $C_m$ [Figs. \ref{fig:figure2}(a) and \ref{fig:figure2}(b)], we  first determine the impedances of the  ${100~\rm M \Omega}$ resistor, the contact pads,  and the lock-in inputs at $f_o=160~\rm Hz$. We  ignore the imaginary component of the $100$ $\rm M \Omega$ resistor. The input impedance $Z_{in}$ of the lock-in amplifier can be modeled  \cite{foster1996whole} as a resistor, $R_{in}={10~\rm M \Omega}$, in parallel with a capacitor, $C_{in}={25~\rm pF}$, as shown in Fig. \ref{fig:figure2}(b). This gives an equivalent impedance of $Z_{in}\approx 9.40-j2.36~\rm M\Omega$ at $f_o=160~\rm Hz$. By comparing the results of  two-probe measurements to four-probe measurements, we estimate each contact impedance to be $Z_c \approx 60-j60~\rm k\Omega$ and thus negligible.  We then calculate $I$ from the rms lock-in reference voltage value $V_s$ using Ohm's Law in the simplified circuit  shown in Fig. \ref{fig:figure2}(b). Here, $I$ and $I_{in}$ are the currents that flow through the microfluidic resistor and the amplifier input circuit, respectively. Using  these properly determined current values, we  then find the values of $R_m$ and $C_m$ by linear fitting, i.e., Ohm's Law, as shown in Fig. \ref{fig:figure3}(a).  These fits yield   $R_m \approx 3.3~\rm M \Omega$ and $3.2~\rm M \Omega$  for live and dead cells, respectively, and  $C_m \approx 0.2~\rm nF$,  resulting in  $Z_m \approx 2.30-j1.50~\rm M\Omega$ for both cases. This  capacitance value is consistent with the parasitic capacitance of the wiring and the cables. We emphasize that the value of $R_m$ depends on the number $N$ of  cells trapped in the device \cite{yang2020all}, and $N\approx 300\pm 50$ for both measurements in Fig. \ref{fig:figure3}(a). The $N$ value, however, will be varied in some measurements and its effects will be deconvoluted from the measurements, as described below.

\subsubsection{Johnson-Nyquist noise}

Now, we turn to a typical noise data trace and discuss the general features of noise. Figure \ref{fig:figure3}(b) shows the time-domain voltage fluctuations, $v(t)$, measured across the microfluidic resistor filled with  roughly 300 cells in LB for a bias current of $I\approx 7.74$ nA  in a (noise) bandwidth of $\Delta f \approx 10 \rm~Hz$. Within the 120 s data trace, the bacteria do not divide or move into and out of the microchannels. The rms value of the voltage noise can be found as $1.55~\mu \rm V$ for  live cells (red) and $0.76~\mu \rm V$ for fixed cells (black). The probability distribution of the noise in both cases is nearly Gaussian  [Fig. \ref{fig:figure3}(c)].  

To understand the origin of this noise, we first estimate  the Johnson-Nyquist noise  of the microfluidic resistor and the  input noise of the lock-in amplifier, which together should result in a white thermal spectrum away from the carrier. The  diagram of the  Th{\'e}venin equivalent noise circuit is shown in Fig. \ref{fig:figure2}(c).  Here, $Z_m$  is the  source impedance,  $Z_{in}$ is the amplifier input impedance, and $e_{th}$ is the thermal noise voltage generated by the source impedance with  a PSD of ${4{k_B}T {\Re}{\{Z_m\}}}$ with ${\Re}$ denoting the real part of the complex impedance. We approximate the noise arising from the amplifier as follows. We assume  that the amplifier adds  the equivalent input-referred  voltage noise  ${e_{in}}$  to the thermal noise, as shown in Fig. \ref{fig:figure2}(c). We note that the current noise  of the amplifier is also lumped into ${e_{in}}$.  Then, the measured  voltage noise PSD with respect to the reference nodes $A$ and $B$ in Fig. \ref{fig:figure2}(c) becomes 
\begin{equation}
\begin{split}
    S_V^{(th)}(f,0) & = {\left[ {4{k_B}T {\Re}{\{Z_m\}}} + { {\left< {{e_{in}}^2} \right>} \over {\Delta f }} \right]} { \left| {Z_{in}} \right|^2 \over {\left| Z_m +Z_{in} \right|}^2},  \\
     & = 4{k_B}T R_{n}.
\label{eq:whitepsd} 
\end{split}
\end{equation}
Here, $S_V^{(th)}(f,0)$ indicates that the bias current is zero and $\Delta f$ is the measurement bandwidth. The resistance $R_{n}$ is an equivalent noise resistance representing all the white thermal noise sources in the system. To determine  ${\left< {{e_{in}}^2} \right>}/\Delta f$  of the lock-in amplifier for our sensitivity setting and  source impedance value at 160 Hz, we measure the output noise in separate experiments as a function of  source resistance \cite{mancini2003op}.  These measurements allow us to find the  equivalent input-referred voltage noise  PSD and the equivalent input-referred  current noise  PSD for each amplifier input as approximately $ 1.70 ~\times 10^{-15}~\rm {V^2}/{Hz}$ and $ 2.30~\times 10^{-27}~\rm {A^2}/{Hz}$, respectively. Since the lock-in amplifier is used in the ${V_A}-{V_B}$ mode in our measurements, we calculate the total input-referred noise PSD from the amplifier inputs by adding the noise PSDs from each amplifier input, leading to ${\left< {{e_{in}}^2} \right>}/\Delta f \approx 3.85 \times 10^{-14}~\rm {V^2}/{Hz}$. Finally, by substituting  $Z_m \approx 2.30-j1.50~\rm M\Omega$ (for 300 cells), $Z_{in}\approx 9.40-j2.36~\rm M\Omega$ and ${\left< {{e_{in}}^2} \right>}/\Delta f \approx 3.85 \times 10^{-14}~\rm {V^2}/{Hz}$  into Eq. (\ref{eq:whitepsd}), we find the white noise PSD at the output to be approximately  $S_V^{(th)}(f,0)\approx 4.80~\times 10^{-14}~\rm {V^2}/{Hz}$.  The detectable Johnson-Nyquist voltage noise  within a bandwidth of 10 Hz should therefore be   $\sqrt{4.80\times 10^{-14}~\rm {V^2}/{Hz} \times 10 ~Hz} \approx 0.70~\rm \mu V$. 

Returning to Fig. \ref{fig:figure3}(a),  we realize that  the rms voltage noise values reported above for both live ($1.55~\rm \mu V$) and dead cells ($0.76~\rm \mu V$) are  larger than the Johnson-Nyquist  noise voltage ($0.70~\rm \mu V$). This  qualitatively suggests that  ``excess"  $1/f$ noise must be  dominating for both live and dead bacteria.  What is also remarkable and perhaps unexpected is the enhanced electrical noise  of live cells as compared to  dead cells. The excess $1/f$ noise will  precisely be the topic of our detailed study.

\subsubsection{Excess noise}
\label{sec:excessnoise}

In order to understand the source of the excess voltage noise in the microfluidic resistor, we turn to equilibrium resistance noise.  We first derive the  dependence of the noise power on   bias current. As shown in Fig. \ref{fig:figure2}(d), the applied bias current $I$ is divided into two, ${I_R}$ through  $R_m$ and  ${I_C}$ through  $C_m$, so that $I={I_R}+{I_C}$. For $I_R$, we find $I_R = {I \over (1+j\omega_o {R_m}{C_m}) }$, where ${\omega_o \over 2\pi}= f_o=160~\rm Hz$. In Fig. \ref{fig:figure2}(d), the resistance noise is modeled as being generated by a  time-dependent fluctuating resistance  $r(t)$ in series with $R_m$ \cite{scofield1987ac}. Under bias  current $I$, the resistance fluctuations are turned into voltage fluctuations via Ohm's Law. The  PSD of the excess voltage noise from the resistance fluctuations is $ {{I^2 S_R(f)} \over {1+{\omega_o}^2 {R_m}^2 {C_m}^2}}$, where $S_R(f)$ is the PSD of the resistance fluctuations in units of $\rm \Omega^2 /Hz$.  Thus, the PSD of the current-dependent  excess voltage noise  measured between the nodes $A$ and $B$ in Fig. \ref{fig:figure2}(d) can be expressed as 
\begin{equation}
\begin{split}
S_V^{(ex)}(f,I) & = I^2 S_R(f) { \left[ {  \left|{{Z_{in}}} \right|^2 } \over \left(1+{\omega_o}^2 {R_m}^2 {C_m}^2 \right)^2 {\left| Z_m +Z_{in} \right|}^2 \right] }, \\
& =  I^2 S_R(f) {\cal C}.   
\end{split}
\label{eq:exPSD}
\end{equation}
Here, the factor  ${\cal C}$ is a dimensionless coefficient that quantifies how the noise generated in the microfluidic resistor is attenuated at the output.  In our experiments, the factor ${\cal C}$ is assumed to be only a function of $R_m$, since $Z_{in}$ and the  capacitance $C_m$ coming mostly from the wiring stay constant. The value of $R_m$  changes with the number $N$ of bacteria in the microchannel and/or the resistivity of the different electrolytes.  Using the circuit parameters given above, i.e., $C_m$, $Z_{in}$, and $\omega_o$,  we can readily determine  ${\cal C}$ as a function of $R_m$ as shown in Appendix \ref{app:dimensionless}.

In the experiments, we measure the total voltage noise PSD as 
\begin{equation}
\begin{split}
S_V^{(tot)}(f) & =  S_V^{(ex)}(f,I) + S_V^{(th)}(f,0),\\
& =  {\cal C}(R_m) I^2 S_R(f) +4k_BTR_n. 
\end{split}
 \label{eq:totPSD}
\end{equation}
When comparing measurements with different  $R_m$ values, it is thus necessary to deconvolute the effects of $R_m$ from  the measured noise for consistency. This should be the case, for instance, when comparing data taken in different electrolytes or when the noise power is measured as a function of number $N$ of bacteria   in the microchannels. In previous work, we have established that the $R_m$ value depends on $N$ approximately linearly as $R_m(N) \approx 2.5~{\rm M\Omega} + N\times 2.5~{\rm k\Omega}$ for \textit{K. pneumoniae} and as $R_m(N) \approx 2.5~{\rm M\Omega} + N\times 3.5~{\rm k\Omega}$ for \textit{S. saprophyticus} \cite{yang2020all}. To compare noise measurements, one should  first subtract from a given $S_V^{(tot)}(f)$ data the thermal noise contribution  $S_V^{(th)}(f,0)$. Then by using the ${\cal C}(R_m)$ corresponding to the $R_m$ value of the microfluidic resistor, one can  obtain the PSD of the resistance fluctuations  as
\begin{equation}
  S_R(f)={ S_V^{(tot)}(f)-S_V^{(th)}(f,0) \over {I^2{\cal C}(R_m)} }. 
 \label{eq:excessPSD}
\end{equation}
The so-called normalized PSD $S(f)$ can then be calculated as  \cite{scofield19851}
\begin{equation}
\begin{split}
 S(f) & = {S_R(f) \over {R_m}^2},\\
 &={ S_V^{(tot)}(f)-S_V^{(th)}(f,0) \over  {\cal C}(R_m) I^2 {R_m}^2}.   
 \end{split}
 \label{eq:normalizedPSD}
\end{equation}

\subsection{Data analysis}
\label{sec:dataanalysis}

%*************************************Figure************************
%%%%%%%%%%%%%%%%%%%%%%%%%%%%%%%%%%%%%%%%%%%%%%%%%%%%%%%%%%%%%%%%%
%%%%%%%%%%%%%%%%%%%%%%%%%%%%%%%%%%%%%%%%%%%%%%%%%%%%%%%%%%%%%%%%%
\begin{figure*}
    \begin{center}
    \includegraphics[width=6.75in]{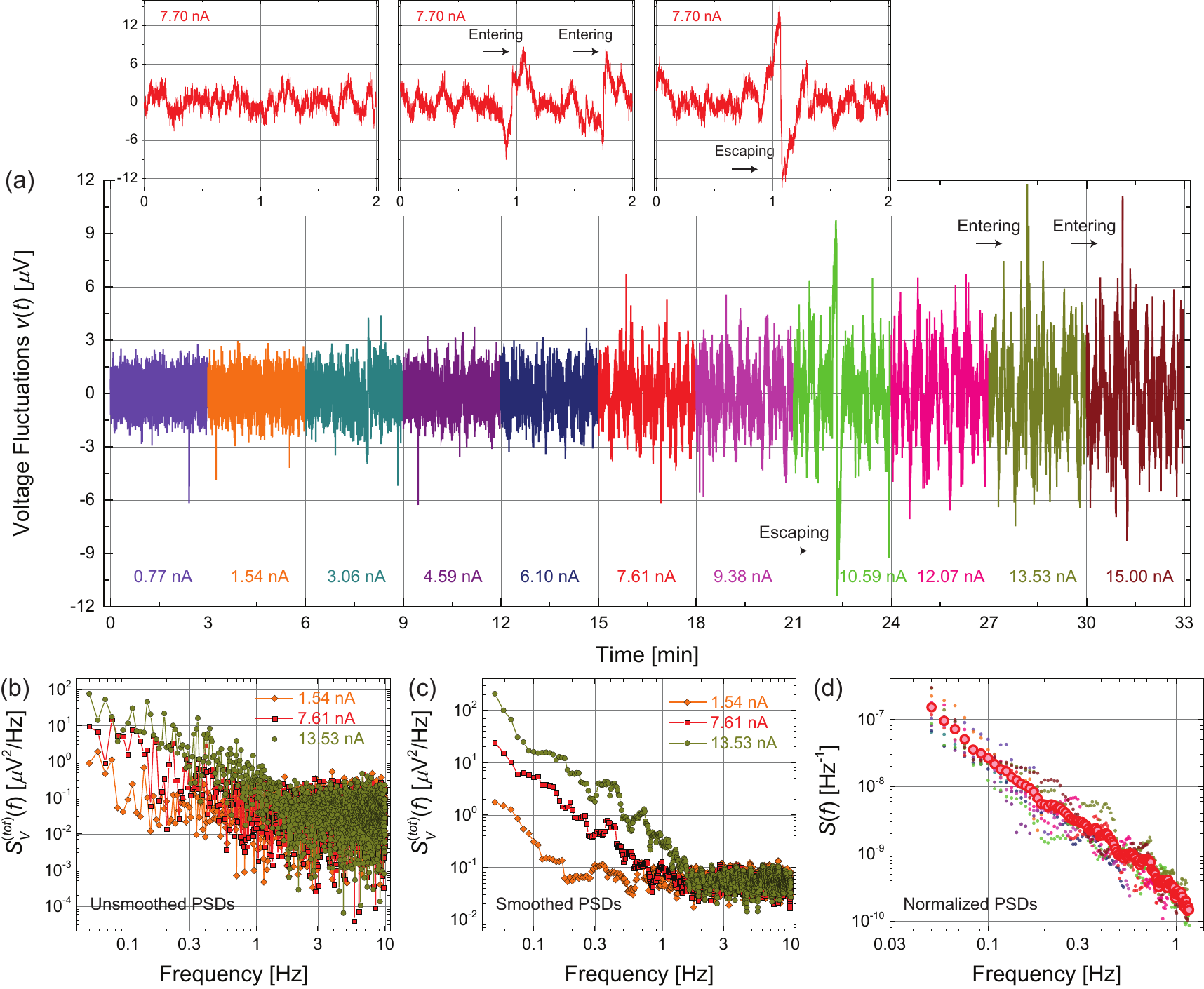} \\ %place holder
\caption{(a) Representative time domain $v(t)$  traces from measurements on live \textit{K. pneumoniae}  in LB at $37^{\circ}$C under different bias currents $0.77~{\rm nA} \leq  I \leq 15.00~{\rm nA}$ with $N\approx 300 \pm 50$ and $\Delta p \approx 0.6~\rm kPa$.  The rms values of $I$ are as indicated in the figure. Upper insets show examples of   $v(t)$ measured at $I\approx 7.70~\rm nA$, with no artifacts (left), with two bacteria entering  the microchannel (center), and with one bacterium escaping from the microchannel (right)---as observed in microscope images. The arrows indicate the instants when the bacteria enter and escape. (b) Representative PSDs  $S_V^{(tot)}(f)$ of $v(t)$ for three different $I$ values before   and (c) after smoothing. (d) Normalized PSDs $S(f)$ of  excess noise. The data show $S(f)$ for each $I$ (small symbols) and average of $S(f)$   over  $I$ (large symbols).}
    \label{fig:figure4}
    \end{center}
\end{figure*}
%%%%%%%%%%%%%%%%%%%%%%%%%%%%%%%%%%%%%%%%%%%%%%%%%%%%%%%%%%%%%%%%%
%%%%%%%%%%%%%%%%%%%%%%%%%%%%%%%%%%%%%%%%%%%%%%%%%%%%%%%%%%%%%%%%%%%

\subsubsection{Basic steps}
\label{sec:basicsteps}

To recapitulate, we measure the voltage noise,  $v(t)$, in time domain as a function of the bias current $I$. We now describe how these data are processed. Figure \ref{fig:figure4}(a) shows representative data traces from our measurements of live \textit{K. pneumoniae}  in LB under different bias currents $I$, with $I$ as indicated in the figure.  The data are first numerically filtered such that the remaining fluctuations are in the frequency range of  $0.05~\rm Hz $ $\leq f$ $\leq 10~\rm Hz$. As seen in the data traces in Fig. \ref{fig:figure4}(a), we  measure  $v(t)$  over a period of 3 min for each applied $I$.   These 3 min traces are thus long enough  that various artifacts can be removed consistently but short enough that bacteria  do not grow substantially and divide. The artifacts arise because bacteria may randomly enter into or escape from the microchannels; these excursions by bacteria can generate spikes  [e.g., Fig. \ref{fig:figure4}(a), center  and right insets]. Although rare, these artifacts can change the characteristics of the noise data. We thus remove these spikes from the measured $v(t)$ traces before we calculate the frequency domain PSDs.  In summary, we  use a 2-min long portion of the  $v(t)$ data  to calculate the PSDs, $S_V^{(tot)}(f)$.   The PSD is calculated by taking a fast Fourier transform (FFT) of the autocorrelation function of a 2-min long $v(t)$ data trace. The PSDs  [Fig. \ref{fig:figure4}(b)] are then smoothed [Fig. \ref{fig:figure4}(c)] using an eight-point moving average \cite{smeets2008noise, heerema20151, siwy2002origin}. To analyze the asymptotic low-frequency behavior of the excess noise, we calculate the normalized PSD, $S(f)$, of the excess noise. For this, we first determine $S(f)$ for each  $I$  using Eq. (\ref{eq:normalizedPSD}).  We then  find the average value  of  $S(f)$ \cite{verleg1998resistance, fleetwood20151}.  The small symbols in Fig. \ref{fig:figure4}(d) show $S(f)$ calculated from $S_V^{(tot)}(f)$  at different  $I$ values, and the large symbols show the average $S(f)$.

\subsubsection{Averaged excess noise}
Finally, we also analyze the dependence of the excess voltage noise on $I$, $\Delta p$, and $N$ below. To this end, we select a frequency range, where  $S_V^{(tot)}(f)$ is significantly above the white (thermal) noise level and determine the PSD of the excess noise from Eq. (\ref{eq:totPSD}), i.e., $S_V^{(ex)}(f)=S_V^{(tot)}(f)-S_V^{(th)}(f,0)$. Here, the PSD of the thermal noise, $S_V^{(th)}(f,0)$, is assumed to be frequency independent. We then average the excess noise over this frequency range as \cite{pal2009resistance, hoogerheide2009probing}
\begin{equation}
{\overline{S_V^{(ex)}(f)}}  = {1\over {f_1-f_2}}{\int_{f_1}^{f_2} S_V^{(ex)}(f) df}.
\label{eq:avednPSD}
\end{equation}
The frequency band we use for these averages is $0.05-0.2~\rm Hz$. 

\subsection{Control experiments}

In a number of additional measurements, we have  characterized the noise in the microchannels  without any bacteria. In these experiments, the microchannels are filled with just the electrolyte solutions, LB and PBS. Some experiments are repeated at $23^{\circ}$C.  The measurements of the
 electrolytes without bacteria are performed with fresh buffers. The experimental approach and data analysis steps are identical to those above.

%*************************************Results************************
\section{Results}
\label{sec:results}

%*************************************Figure************************
%%%%%%%%%%%%%%%%%%%%%%%%%%%%%%%%%%%%%%%%%%%%%%%%%%%%%%%%%%%%%%%%%
%%%%%%%%%%%%%%%%%%%%%%%%%%%%%%%%%%%%%%%%%%%%%%%%%%%%%%%%%%%%%%%%%

\begin{figure*}[ht]
\centering
\includegraphics[width=4.5in]{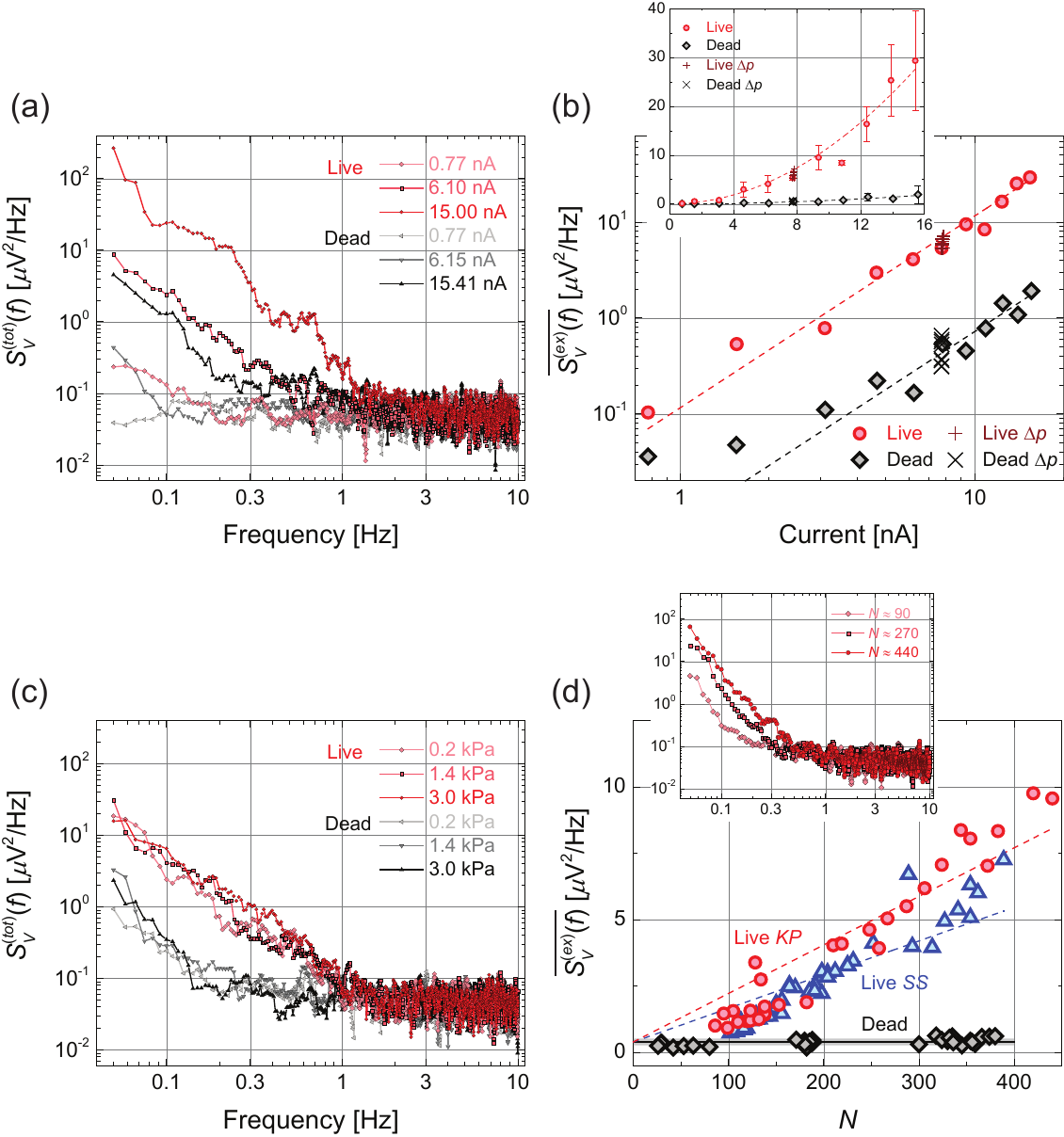}% Here is how to import EPS art
\caption{ (a) Voltage noise PSDs $S_V^{(tot)}(f)$  for different bias $I$  for live and dead (\textit{K. pneumoniae}) cells. There are approximately $300 \pm 50$ cells in both devices, and $R_m$ values are similar. (b) Averaged PSD $\overline{S_V^{(ex)}(f)}$ of the excess voltage noise  as a function of $I$. Each data point is the average from three independent experiments with nominally identical devices that have $300 \pm 50$ cells and similar $R_m$ values. The dashed lines are quadratic fits of the form $\overline{S_V^{(ex)}(f)}=\Gamma I^2$, with  $\Gamma=1.16\times {10}^{5}~\rm \Omega^2/Hz$  (live) and $7.21\times {10}^{3}~\rm \Omega^2/Hz$ (dead). The symbols $+$ and $\times$  respectively show $\overline{S_V^{(ex)}(f)}$ for live and dead cells at different  $\Delta p$ ranging from 0.2 kPa to 3.0 kPa with an increment of 0.4 kPa under $I\approx 7.75 ~\rm nA$. Inset shows the same data   on a linear scale;  error bars show single standard deviations. (c) Representative $S_V^{(tot)}(f)$ for live and dead cells at three different $\Delta p$ values, with $I\approx 7.75 ~\rm nA$. (d) Averaged PSD $\overline{S_V^{(ex)}(f)}$  as a function of  number $N$ of cells  in the microchannels. The bias current is $I\approx 7.75 ~\rm nA$ for all. Each data set consists of data points from  at least three independent experiments. The line parallel to the $x$-axis corresponds  to the excess noise power within the $0.2~\rm Hz$ bandwidth for dead cells. The dashed lines are linear fits with  the $y$ intercepts fixed to the   excess noise value for the dead cells.  Inset shows three representative $S_V^{(tot)}(f)$ curves for live  \textit{K. pneumoniae} cells with different $N$ taken using the same bias $I\approx 7.78 ~\rm nA$.}
\label{fig:figure5}
\end{figure*}
%%%%%%%%%%%%%%%%%%%%%%%%%%%%%%%%%%%%%%%%%%%%%%%%%%%%%%%%%%%%%%%%%
%%%%%%%%%%%%%%%%%%%%%%%%%%%%%%%%%%%%%%%%%%%%%%%%%%%%%%%%%%%%%%%%%

We first make observations on the   frequency domain characteristics of the measured noise. These observations lead us to the conclusion that the noise is due to equilibrium resistance fluctuations. We then show the scaling behavior of the measured noise by obtaining its normalized PSD $S(f)$.

\subsection{Noise PSD as a function of different parameters}

\subsubsection{Bias current}

Figure \ref{fig:figure5}(a) shows the PSDs    of the voltage fluctuations, $S_V^{(tot)}(f)$, measured in microchannels filled with $N \approx 300 \pm 50$ live  and dead  cells at different  bias current values. All the PSDs are obtained  from time domain data, such as those in Fig. \ref{fig:figure4} using the steps described in Section \ref{sec:dataanalysis}. As noted above, the time domain data traces   do not contain any voltage spikes due to large perturbations, such as bacterial divisions or displacements.  Several observations are noteworthy in Fig. \ref{fig:figure5}(a). For each $I$, the PSD exhibits a well-defined  frequency $f_w$, where the behavior of the curve changes. For $f>f_w$, the  spectrum is white and independent of $I$; the experimentally measured white noise PSD of $5.25 \times 10^{-14}~\rm {V^2}/{Hz}$ is consistent with the amplifier input noise combined with the Johnson-Nyquist noise of the microfluidic device, which we have estimated  above in Section \ref{sec:electricalmeasurments} to be $4.80 \times 10^{-14}~\rm {V^2}/{Hz}$.   For  $f<f_w$, the PSDs increase   as frequency decreases, showing typical $1/f$ excess noise characteristics \cite{hooge19941} for both live and dead cells. The value of $f_w$ also appears to increase with $I$.

%*************************************Figure************************
%%%%%%%%%%%%%%%%%%%%%%%%%%%%%%%%%%%%%%%%%%%%%%%%%%%%%%%%%%%%%%%%%
%%%%%%%%%%%%%%%%%%%%%%%%%%%%%%%%%%%%%%%%%%%%%%%%%%%%%%%%%%%%%%%%%
\begin{figure}[!ht]
\centering
\includegraphics[width=3.375in]{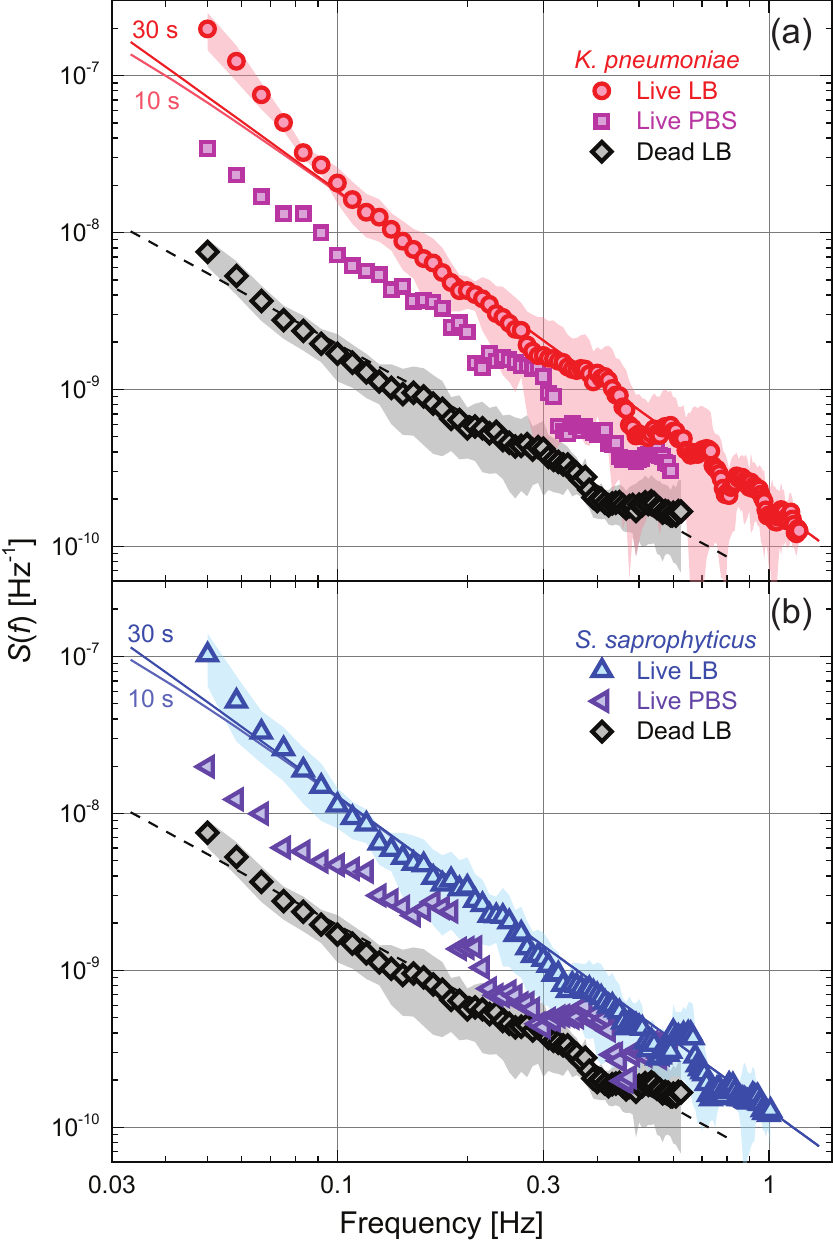}% Here is how to import EPS art
\caption{Collapse and scaling of the noise data. Normalized PSDs, $S(f)$, of  resistance fluctuations for live \textit{K. pneumoniae} (a) and \textit{S. saprophyticus} (b)  in LB and PBS. The black data trace  in each plot shows  the average $S(f)$ for dead \textit{K. pneumoniae} in LB and dead \textit{S. saprophyticus} in LB.  The shaded regions show the error (single standard deviations). Each  data trace on live and dead cells in LB are obtained from  three independent experiments. The data trace on live cells in PBS is from a single experiment. The black dashed lines are $S(f)={{B} / {f^ \beta}}$, with $B=6.15 \times {10}^{-11}$ and $\beta=1.50$.  The solid lines are fits to  $S(f)={{A {\tau}^2 } \over {1 + 4{\pi ^2}{f^2}{{\tau} ^2}}}$ using  $\tau= {10~\rm s}$ and ${30~\rm s}$ for \textit{K. pneumoniae} (red) and \textit{S. saprophyticus} (blue); the $A$ values for both time constants for \textit{K. pneumoniae} and \textit{S. saprophyticus} are approximately $7.30 \times {10}^{-9}$ and  $5.10 \times {10}^{-9}$, respectively.}
\label{fig:figure6} 
\end{figure}
%%%%%%%%%%%%%%%%%%%%%%%%%%%%%%%%%%%%%%%%%%%%%%%%%%%%%%%%%%%%%%%%%
%%%%%%%%%%%%%%%%%%%%%%%%%%%%%%%%%%%%%%%%%%%%%%%%%%%%%%%%%%%%%%%%%

We now turn to the dependence of the excess noise on $I$. Here, we compare  ${\overline{S_V^{(ex)}(f)}}$ values, as defined in Eq. (\ref{eq:avednPSD}) above,  for different $I$ \cite{pal2009resistance, hoogerheide2009probing}.  Figure \ref{fig:figure5}(b) shows  $\overline{S_V^{(ex)}(f)}$ as a function of $I$. For live cells, the $\overline{S_V^{(ex)}(f)}$ data can be fitted to a quadratic function  as $\overline{S_V^{(ex)}(f)} \approx \Gamma_l I^2$, with $\Gamma_l \approx 1.15 \times 10^5~\rm {\Omega^2}/{Hz}$. The  noise of dead bacteria in Fig. \ref{fig:figure5}(b)  can also be fitted to a quadratic function,  $\Gamma_d I^2$---at least for  the high current region of the data with  $\Gamma_d \approx 7.20 \times 10^3~\rm {\Omega^2}/{Hz}$.  The dashed lines in Fig. \ref{fig:figure5}(b) show the quadratic fits.  It is important to emphasize that all the data here are consistently taken on $N \approx 300 \pm 50$ cells, and all $R_m$ values  remain in the range $R_m \approx 3.10 \pm 0.40~\rm M\Omega$.

Two important conclusions can be made based on the data and fits in Fig. \ref{fig:figure5}(b). The $I^2$ dependence of the excess noise suggests that, in both cases, the  noise is induced by ``equilibrium" resistance fluctuations \cite{voss1976flicker, dutta1981low, verleg1998resistance, tasserit2010pink, wen2017generalized}. Live cells exhibit a significantly higher  amplitude than dead ones with $\Gamma_l \gg \Gamma_d$, indicating that noise  generation is linked to  bacterial metabolism. The origin of the noise from  dead cells is not entirely clear and will be further addressed in Section \ref{sec:excessnoise} below. 

\subsubsection{Hydrodynamic pressure}

Fig. \ref{fig:figure5}(c) shows $S_V^{(tot)}(f)$  for different applied pressures $\Delta p$ across the channel at  a fixed bias of  $I\approx 7.75 ~\rm nA$.  The measured noise remains independent of $\Delta p$, and hence the bulk flow velocity, as $\Delta p$ is varied by more than an order of magnitude. We further note that $\overline{S_V^{(ex)}(f)}$ at $I\approx 7.75 ~\rm nA$ exhibits no dependence on $\Delta p$ [Fig. \ref{fig:figure5}(b)].

During the experiments,  we establish a net flow from the inlet to the outlet. The drag force due to this steady flow  has a stabilizing effect on the system, with the bacteria snugly jammed in the microchannel  toward the nanoconstriction. The electrokinetic and other flow forces in the system do not move the bacteria due to the presence of this net steady flow from the inlet toward the nanoscale constriction. Appendix \ref{app:perturbations} provides more details on the control experiments where we measure the hydrodynamic and electrokinetic forces on the bacteria in our system.

\subsubsection{Number of cells}

We next demonstrate how the number $N$ of cells in the microchannel affects the measured noise characteristics. We increase the number of bacteria in the microchannels via growth for live cells and via trapping from flow for dead cells. We  measure $v(t)$ at fixed $I\approx 7.75~\rm nA$ as a function of $N$. The inset of Fig. \ref{fig:figure5}(d) shows  $S_V^{(tot)}(f)$ of live cells  for three different $N$ values, with the PSD increasing with $N$ in the low-frequency region. The main plot in Fig. \ref{fig:figure5}(d) shows  $\overline{S_V^{(ex)}(f)}$  as a function of $N$ obtained from many different data traces such as those in the inset. While $\overline{S_V^{(ex)}(f)}$ increases monotonically with $N$ for live cells,  it remains at a constant value (solid line) for dead cells.  The  increase of  $\overline{S_V^{(ex)}(f)}$ with $N$ for live cells indicates that the noise powers from individual cells are additive. The   noise power of \textit{K. pneumoniae}  appears to be  slightly larger than that of \textit{S. saprophyticus}. We note that, since $R_m$ increases with $N$, the fraction of the noise power that is coupled to the amplifier changes with $N$.  Thus, the data only provide a qualitative picture.

%\subsubsection{Control Experiments: Electrokinetic Forces on Bacteria}

\subsection{Normalized PSD of the excess noise}
\label{sec:excessnoise}

\subsubsection{Excess noise of live bacteria}

The $I^2$ dependence observed in Fig. \ref{fig:figure5}(b) suggests that the  noise is due to equilibrium resistance fluctuations.  To  analyze the asymptotic low-frequency behavior of the excess noise, we return to Eq. (\ref{eq:normalizedPSD}) and calculate the normalized PSD, $S(f)$, of the excess noise. Normalizing the data  this way removes  the contribution of the thermal noise and all the ${I}$ and $R_m$ dependences.  The data with different   bias currents should then collapse onto a single curve  \cite{hooge19691}.  $S(f)$ of  different systems, e.g., dead and live bacteria,  can then be directly compared.

The collapsed   $S(f)$ data are shown in Fig. \ref{fig:figure6}(a) for \textit{K. pneumoniae} and in Fig. \ref{fig:figure6}(b) for \textit{S. saprophyticus}, both in LB and PBS. For live bacteria in LB,   $S(f) \propto {{{A}} {{f^{-2}}}}$. The noise of live cells in PBS is noticeably lower than that in LB. In all these measurements, $N \approx 300 \pm 50$.

\subsubsection{Excess noise of dead bacteria and electrolytes}

The black data traces  in Figs. \ref{fig:figure6}(a) and \ref{fig:figure6}(b) show  the average $S(f)$ for dead \textit{K. pneumoniae} and dead \textit{S. saprophyticus}, respectively, with the shaded regions corresponding to the error (single standard deviations). For dead bacteria, $S(f) \propto  {{{B}}  {{f^{-3/2} }}}$.

\begin{figure}
\centering
    \includegraphics[width=3.375in]{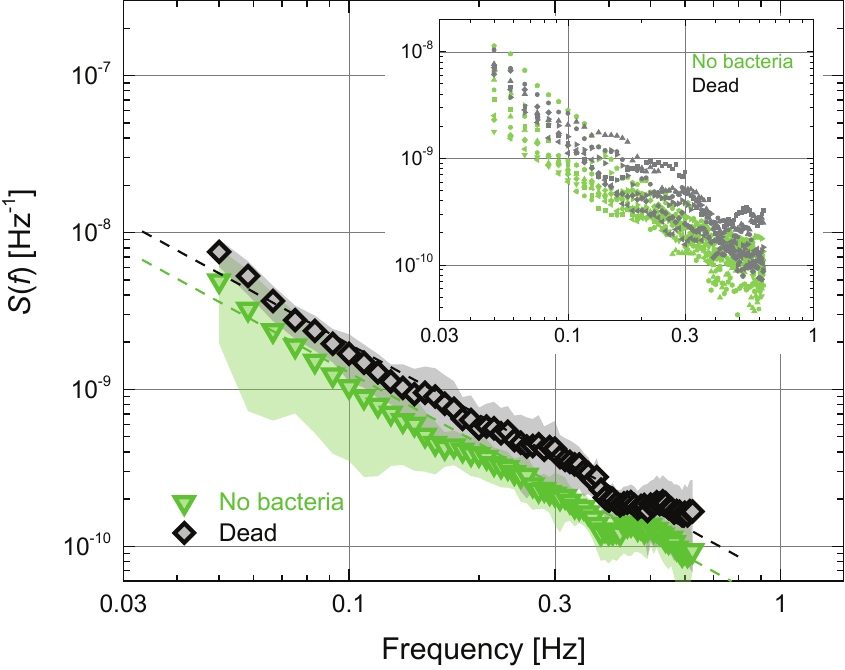}  %place holder
    \caption{Normalized PSD $S(f)$ of the resistance fluctuations for electrolytes containing no bacteria and dead bacteria. The large symbols show the average values of the experiments with just the electrolyte (green) and with dead bacteria (black); the shaded regions show the error (single standard deviations); the dashed lines are not fits but show the asymptotic behavior and correspond to ${S(f)}={B/f^{\beta}}$, with $B = 4.05 \times {10}^{-11}$ and $\beta = 1.50$ for no bacteria (green) and  $B = 6.15 \times {10}^{-11}$ and $\beta = 1.50$ for dead bacteria (black). The inset shows the data sets from independent experiments; the  green  symbols correspond to data from nine independent experiments, three with PBS at $37^{\circ}$C, three with LB at $37^{\circ}$C, and three with  LB at $23^{\circ}$C; the  gray symbols show data from six independent experiments three with dead \textit{K. pneumoniae} and three with dead \textit{S. saprophyticus}, both in LB at $37^{\circ}$C.}
    \label{fig:nodead}
\end{figure}

Microchannels filled with electrolytes (either LB or PBS) with no  cells at both $37^{\circ}$C and $23^{\circ}$C show the same noise characteristics and levels  as  microchannels filled with dead bacteria. This can be seen clearly in the collapsed average $S(f)$ data  in Fig. \ref{fig:nodead}.  $S(f)$ curves in Fig. \ref{fig:nodead} for no bacteria and dead bacteria also follow the same scaling behavior. In order to obtain these collapse plots, we follow the steps described above in Section \ref{sec:basicsteps} and  use  ${\cal C}(R_m)$ to correct the measured noise  in different electrolytes. This is because the electrical conductance values, and hence the $R_m$ values, for these devices differ significantly: at $37^{\circ}$C, the  $R_m$ values for devices filled with LB and PBS are $2.5~\rm M \Omega$ and $1.7~\rm M \Omega$, respectively; at room temperature ($23^{\circ}$C), the  $R_m$ value for LB  is $3.0~\rm M \Omega$; the electrical conductivity of PBS at $37^{\circ}$C, LB at $37^{\circ}$C and LB at $23^{\circ}$C are approximately  1.47 S/m, 1.00 S/m, and 0.83 S/m, respectively. These conductivity values agree with those reported in the literature \cite{binette2004tetrapolar,prieto2016monitoring,li2018differentiation}.

In summary, the presence or absence of dead cells in the electrolyte does not change the background noise appreciably; nor does interchanging  LB with PBS or decreasing the temperature to $23^{\circ}$C. More details are available in the Supplemental Material \cite{yang2021supp}. It may therefore be justifiable to collapse all the background data onto a single curve.  All these suggest that the background noise  is  due to an intrinsic bulk process in the electrolyte---although we cannot conclusively rule out other possibilities, such as contact noise \cite{hassibi2004comprehensive} or extrinsic noise. 

%*************************************Discussion************************
\section{Discussion}
\label{sec:discussion}

We first systematically look at possible artifacts, concomitants, and perturbations that may give rise to the observed noise. These include extrinsic noise, measurement artifacts, bacterial movements, flow forces, electrokinetic forces, electrical perturbations, and Joule heating. For each case, we provide estimates or experimental evidence that suggests that the noise does not come from this particular source. (Some of the details are presented in the Appendices.) We then return to our hypothesis that bacterial metabolism is responsible for the observed noise and provide some theoretical support to our experimental observations.

\subsection{Possible artifacts}

We first consider possible artifacts and perturbations that might give rise to the observed excess noise. The fact that we can directly compare results on live and dead cells has  convinced us that our measurements are not dominated by extrinsic noise. Our microchannel device is, in principle, capable of transducing  bacterial movements into resistance fluctuations \cite{kara2018microfluidic}. We can rule out any inherent rigid body movements of live bacteria (e.g., wiggling) as a noise source because our  bacteria  are highly non-motile. The steady flow and the electrokinetic flows due to applied fields in the microchannel can also excite and sustain rigid body movements \cite{lissandrello2015noisy}. However, this cannot be the source of the observed noise simply  because  fixed  cells would  show the same level of noise [e.g., Fig. \ref{fig:figure2}]. Detailed measurements described in Appendix \ref{app:perturbations} indicate that  electrokinetic forces are negligible compared to the steady drag force of the pressure-driven flow, which pushes the bacteria toward the nanoconstriction and jams them. We also investigate a possible  temperature increase  due to Joule heating and conclude that  the increase is negligible. An applied electric field, i.e., the bias, can  perturb the bacterial  membrane and its proteins \cite{tsong1990electrical,xie1990study,stratford2019electrically}, changing the influx and efflux rates. Our electric field strength is too small to induce any electrical perturbations, and bacteria appear to grow normally in our microchannels during our measurements.  We therefore conclude that the observed  noise is rooted in bacterial metabolism.

\subsection{Nanomechanical fluctuations of bacteria}

Active random oscillations and movements are  commonly observed in many microorganisms and cells \cite{ben2011effective,turlier2016equilibrium,biswas2017mapping,nelson2017vibrational,wu2016quantification,singh2019studying, manneville2001active, willaert2020single}. Due to active biochemical processes in the cytoplasm and cell membrane \cite{longo2013rapid, lissandrello2014nanomechanical}, highly  non-motile bacteria  are also expected to exhibit random nanomechanical oscillations. During these nanomechanical oscillations, the bacteria go through quasi periodic deformations, which in principle could get converted to electrical noise by our microchannel transducer \cite{kara2018microfluidic}.

In order to get an order of magnitude estimate of nanomechanical fluctuations, we convert the observed electrical noise power into an active effective temperature $T_{eff}$ for the bacteria. To this end, we first numerically calculate the rms thermal amplitude, i.e., nanomechanical thermal equilibrium noise,  of a bacterium in several of its eigenmodes; we  then convert the nanomechanical noise  into  electrical resistance noise by considering the change in the geometric cross-section of the microchannel due to the random deformations of $300$ bacteria. Based on the responsivity of our microchannel transducer, the random thermal oscillations  of 300 bacteria at the equilibrium temperature of $310~\rm K$ ($37^{\circ}$C) result in resistance noise with an rms  value of $\sim 0.8~\Omega$.  Experimentally measured rms resistance noise, on the other hand, is $\sim 200~\Omega$. To match the observed noise levels, we estimate that a bacterium ought to attain an effective temperature of $T_{eff}\sim 10^7~\rm K$. While an active system such as a bacterium should have $T_{eff} > 310~\rm K$ \cite{ben2011effective,turlier2016equilibrium,biswas2017mapping,manneville2001active},  this level of activity seems unreasonable. Nanomechanical motion of bacteria is thus an unlikely source for our observations. Details of the simulations, the assumptions made, and a  thorough discussion of the results are provided in Appendix \ref{app:perturbations}. 

\subsection{Charge noise model}

Our hypothesis is that the source of the fluctuations is electrical.   The electrolyte-filled microchannels can be considered as  one-dimensional conductors in which the primary charge carriers are  $\rm Na^+$ and  $\rm Cl^-$ ions. Each microchannel is  assumed to be long and uniformly filled with  bacteria (i.e., $N\approx 300$). Monovalent ions, such as $\rm K^+$, $\rm Na^+$ and $\rm Cl^-$, move into and out of the bacteria randomly; this  causes fluctuations in the number of  charged carriers within the conductors and hence resistance noise. 

We first construct an electrical noise model of a single cell by  assuming that the   flux of a certain ion $\rm X$ into or out of the cell  is  proportional to the number of ions within the cell. For instance, if there are excess $\rm Na^+$ ions in the cell, $\rm Na^+$ ions will be transported  out of the cell  and \textit{vice versa}.  This is expected because the excess $\rm Na^+$ in the cell will change the Nernst potential for $\rm Na^+$ and  activate the ion channel conduction in one direction. Thus, for any given ion $\rm X$,
\begin{equation}
    {{d { n_{\rm X}} } \over {dt}} =  -{\phi}({ n_{\rm X}}), 
\end{equation}
where ${ n_{\rm X}}$ is the total number of intracellular $\rm X$ ions and ${\phi}$ is the rate of  transport of $\rm X$ ions through the cell membrane as a function of $n_{\rm X}$. The minus sign indicates that when intracellular value of ${ n_{\rm X}}$ is decreasing the ${ n_{\rm X}}$ flux is positive.  Expanding around the equilibrium value  $n_{\rm X} ={\bar n_{\rm X}} + \Delta n_{\rm X}$, we write
\begin{equation}
    {{d {\Delta n_{\rm X}} } \over {dt}} =  - {{\Delta n_{\rm X}} \over \tau_{\rm X} } + \xi_{\rm X},
\end{equation}
with $\xi_{\rm X}$ being a white noise term. Here, $\phi({\bar n_{\rm X}})= 0$, and ${\tau _{\rm X}} = {\left( {{{\left. {{{\partial \phi} \over {\partial {n_{\rm X}}}}} \right|}_{{{\bar n}_{\rm X}}}}} \right)^{ - 1}}$ can be regarded as a lifetime for ions or the relaxation time for an ionic perturbation within the cell.  From this, we find \cite{van1976noise,  mitin2002generation}
\begin{equation}
    {S_{n_{\rm X}}}(f) = {4{\overline {{{(\Delta n_{\rm X})}^2}} \tau_{\rm X} } \over {1 + 4{\pi ^2}{f^2}{{\tau_{\rm X}} ^2}}}
\end{equation}
for the PSD for the  fluctuations of the number ${ n_{\rm X}}$ of  intracellular ions of type $\rm X$  in a single bacterium.  Since we are interested in estimating an order of magnitude, we assume that the $\tau_{\rm X}$ for different ions are roughly equal and define an overall effective time constant $\tau$ such that $\tau \sim \tau_{\rm X}$.  We also assume that the noise powers generated by   bacteria are additive. Then, the PSD of the fluctuations in the total number of charge carriers within our microfluidic resistor containing many bacteria should be expressible in  the form 
\begin{equation}
{S_{n}}(f) = {4{\overline {{{(\Delta n)}^2}} \tau } \over {1 + 4{\pi ^2}{f^2}{\tau ^2}}},
\end{equation}
where ${\left( {\overline {{{(\Delta n)}^2}} } \right)^{1/2}} = {\Delta n_{rms}}$ is the rms value of the fluctuations in the number of charge carriers. Note that the particle flux at equilibrium is  assumed to be zero---as opposed to the case in generation recombination noise in semiconductors \cite{van1976noise}.

For the equivalent microfluidic resistor $R_m$ (with ten microchannels), the relationship between the PSD of the resistance fluctuations, $S_R (f)$, and the PSD of the carrier number fluctuations, $S_n (f)$, is \cite{mitin2002generation} 
\begin{equation}
    S(f) = {{{S_R}(f)} \over {{{R_m}^2}}} = {{{S_n}(f)} \over {{n}^2}}.
\label{eq:eqpsd}
\end{equation}
Here, $n$ is the total number of charge carriers in the resistor. Under the assumption of spatial uniformity, Eq. (\ref{eq:eqpsd}) holds for a single microchannel (out of the ten) (see Appendix \ref{app:chargenoise}), which has $n\approx 4 \times 10^{10}$ at  the $85$ mM  NaCl concentration of LB.  Thus, there are ${n} \pm { {\Delta n}_{rms}}$ charge carriers within the LB filling the microchannel due to the noise generated by bacteria. 

We do not see the corner frequency in our data and thus  cannot determine $\tau$ from our experiments. We turn to previous  work \cite{kralj2011electrical} for an  approximate value of $\tau$. In this remarkable paper \cite{kralj2011electrical}, the autocorrelation functions of spontaneous electrical blinks from single bacterial cells were shown to decay exponentially  over a period of  $ 10 - 30~\rm s$. This suggests that bacterial membrane potentials and  the intracellular ion concentrations relax with a time constant $ 10~{\rm s} \lesssim \tau \lesssim 30~{\rm s}$. For 10 s and 30 s, we obtain the fits shown in Fig. \ref{fig:figure6} with $\left( \overline {{{(\Delta n)}^2}} \right) \sim 5\times 10^{13}$ for 30 cells at equilibrium yielding ${\Delta n^{(1)}_{rms}} \sim 1.3 \times 10^6$ per cell.  We have  estimated  the other relevant time constants in the system due to flow, drift and diffusion, and found that the diffusion time constant in the system is close to 10 s. It is also noteworthy that the noise data in Fig. \ref{fig:figure6} start to deviate from the $1/f^2$ asymptote below 0.06 Hz, suggesting that another noise process might be dominating  at our lowest frequencies.

\subsection{Estimation of the membrane potential noise}

To estimate the  noise  in the membrane potential due to the charge noise, we use two approaches. 

In the first, we assume that the noise in the transmembrane ionic current results in fluctuations in the total charge  in  close proximity of the membrane, i.e., on the plates of the capacitor in the circuit in Fig. \ref{fig:figure1}(a). Then, $e_n \sim {e\Delta n^{(1)}_{rms} \over C} \approx 3.5~\rm V$, with $\Delta n^{(1)}_{rms} \approx 1.3 \times 10^{6}$, $C\approx 6 \times 10^{-14}~\rm F$ and ${e} \approx 1.60 \times 10^{-19}~{\rm C}$ \cite{benarroch2020microbiologist} with $C$ being the membrane capacitance of the cell. 

In the second, we estimate $e_n$ from fluctuations in the intracellular ion concentrations. We assume that the ions are distributed uniformly inside (and outside) the cell and that only $\rm K^+$, $\rm Na^+$, and $\rm Cl^-$ ions  contribute to the steady-state value of $V_{mem}$. We find the change in the membrane potential with respect to each ion concentration from the   Goldman-Hodgkin-Katz equation \cite{benarroch2020microbiologist} and find the  total rms change by adding each contribution as
%\begin{equation}
%\begin{split}
%    {e_n} & \approx  {{RT} \over F } \left({{{\left(p_{\rm K} \Delta [{\rm K^+}]_i\right)^2} +{\left(p_{\rm Na} \Delta [{\rm Na^+}]_i\right)^2} } \over {\left({p_{\rm K}[{\rm K^+}]_i} +{p_{\rm Na}[{\rm Na^+}]_i}+{p_{\rm Cl}[{\rm Cl^-}]_o}\right)^2}} \right. \\
%    & \left.  + {{\left(p_{\rm Cl} \Delta [{\rm Cl^-}]_i\right)^2} \over {\left({p_{\rm K}[{\rm K^+}]_o} +{p_{\rm Na}[{\rm Na^+}]_o}+{p_{\rm Cl}[{\rm Cl^-}]_i}\right)^2}} \right) ^{1/2},
%\label{eq:eqen}
%\end{split}
%\end{equation}
\begin{widetext}
\begin{equation}
\begin{split}
     {e_n} \approx  {{RT} \over F } {\left({{{\left(p_{\rm K} \Delta [{\rm K^+}]_i\right)^2} +{\left(p_{\rm Na} \Delta [{\rm Na^+}]_i\right)^2} }\over {\left({p_{\rm K}[{\rm K^+}]_i} +{p_{\rm Na}[{\rm Na^+}]_i}+{p_{\rm Cl}[{\rm Cl^-}]_o}\right)^2}} + {{\left(p_{\rm Cl} \Delta [{\rm Cl^-}]_i\right)^2} \over {\left({p_{\rm K}[{\rm K^+}]_o} +{p_{\rm Na}[{\rm Na^+}]_o}+{p_{\rm Cl}[{\rm Cl^-}]_i}\right)^2}} \right)^{1/2}},
\label{eq:eqen}
\end{split}
\end{equation}
\end{widetext}
where $R$ is the universal gas constant, $T$ is the temperature, $F$ is Faraday's constant;  $p_{\rm {X}}$ is the relative membrane permeability, and $[{\rm X}]_i$ and $[{\rm X}]_o$ are the uniform intracellular and extracellular concentrations of the ion. The rms fluctuations in intracellular ion concentration for each ion is estimated to be of order, ${\Delta [{\rm X}]_i} \sim {\Delta n^{(1)}_{rms} \over V_b}\sim 10^{24}~{\rm m^{-3}}$, for a cell with  volume $V_b \sim 10^{-18}~{\rm m^{3}}$ \cite{benarroch2020microbiologist}. The extracellular ion concentrations are assumed to remain constant. By substituting the ${\Delta [{\rm X}]_i}$ values along with the values of $p_{\rm {X}}$, $[{\rm X}]_i$, and $[{\rm X}]_o$ into Eq. (\ref{eq:eqen}), we find $e_n \sim 1.3~\rm mV$. More details are available in Appendix \ref{app:voltagenoise}.

These two extreme values  for the membrane potential fluctuations, i.e., 1.3 mV and 3.5 V, suggest that the fluctuations depend strongly upon how   nonuniformities in  ion concentrations get dissipated within the ``crowded environment" of the cell.  We estimate that an ion would probably diffuse a distance equal to the length scale of a bacterium in $\sim 10$ ms \cite{mcguffee2010diffusion}. This  indicates that a more accurate noise model for the membrane potential and  the electrical fluctuations of the entire bacterium should take into account intracellular diffusion.

%*************************************Conclusions************************
\section{Conclusions}
\label{sec:conclusion}

Our measurements represent  a significant step in understanding the role of fluctuations in bacterial ion homeostasis. By carrying out electrical noise measurements on live and dead bacteria, we have shown evidence that the charge state of live bacteria fluctuates over time. A direct consequence of this observation is that  intracellular ion concentrations and the  membrane potential $V_{mem}$    both have   strongly fluctuating components. It is well established that $V_{mem}$ and intracellular ion concentrations in bacteria  affect a number of cellular processes in a crucial manner. Among these processes are ATP synthesis \cite{mitchell1961coupling}, cell division \cite{strahl2010membrane}, cell motility \cite{guragain2013calcium}, antibiotic resistance \cite{dong2019magnesium}, environmental sensing \cite{bruni2017voltage, anishkin2014feeling} and electrical communication between cells \cite{prindle2015ion, liu2017coupling}.   It is thus reasonable to expect that  $V_{mem}$ and the intracellular ion concentrations should practically remain constant over time, because drastic changes and large fluctuations  in $V_{mem}$ and ion concentrations  would  strongly perturb all the above-mentioned  cellular processes.  More and more experiments, including ours, suggest that  the ionic makeup of the cell, and hence $V_{mem}$, fluctuates strongly. It remains to be seen whether bacteria utilize these fluctuations to its advantage or whether there are built in mechanisms of noise evasion.

We have also established time-resolved electrical resistance measurements as a sensitive tool for studying bacteria. Even lower frequencies and higher sensitivities may be achieved in future studies  by redesigning the microfluidic resistor, e.g., by incorporating a nulling bridge in the design or   and reducing the background fluctuations.  It may also be possible, in principle, to do these experiments in larger channels with many more bacteria---provided that any bacteria movements can be suppressed. Our microfluidic resistor is designed to have a source resistance such that efficient coupling of noise can be achieved to a low-frequency high-impedance amplifier. This requirement can be circumvented, for instance, by performing the measurement at high frequency, where a larger channel with a  source resistance close to $50~\Omega$ can be used.  Finally, understanding the source of the  fluctuations in the electrolyte is also a problem of fundamental relevance. To this end, one may perform experiments with different electrodes and in different electrolytes.

\begin{acknowledgements}

This work was supported by NIH (1R21AI133264-01A1). H. G. acknowledges the Boston University Nanotechnology Innovation Center BUnano Cross-Disciplinary Fellowship. The authors thank V. Yakhot and J. Tien for helpful discussions. 

K. L. E. discloses a potential of conflict of interest, as he is the cofounder of a company, Fluctuate Diagnostics, which aims to commercialize this technology. No potential
conflicts of interest exist for Y. Y. and H. G.

\end{acknowledgements}

\appendix

\section{EXPERIMENTAL DETAILS}
\label{app:experimental}

\subsection{Device design and fabrication}

Our  microfluidic resistor is essentially a continuous  polydimethylsiloxane (PDMS) microfluidic channel that is  bonded onto a glass substrate with pre-patterned metallic electrodes. The device is fabricated using standard soft lithography \cite{yang2020all}. The  microfluidic channel is made up of two structures with different length scales: the large mm-scale channel narrows down, at its center, to ten smaller parallel microchannels ($l\times w\times h \approx 100\times 2 \times 2 ~\mu \rm m^3$). Each microchannel has a nanoconstriction ($l\times w\times h \approx 5 \times 0.8 \times 2 ~\mu \rm m^3$) at one end with  a cross-sectional area close to that of a single bacterium.  More details about the device design and fabrication can be found in \cite{yang2020all}.

\subsection{Bacteria preparation}

Two microorganisms, \textit{Klebsiella pneumoniae} (ATCC 13883) and \textit{Staphylococcus saprophyticus} (ATCC 15305), are purchased from American Type Culture Collection (ATCC, Rockville, MD). \textit{K. pneumoniae} is a Gram-negative, non-motile,  rod-shaped  bacterium that has a length of $2-3 ~\mu \rm m$ and a cross-sectional area of $ 0.8 ~\mu \rm m^2$ \cite{zhang2021measurement}. \textit{S. saprophyticus} is a Gram-positive, non-motile,  spherical bacterium that has a diameter of $ 1 ~\mu \rm m$ \cite{monteiro2015cell}. We use Luria-Bertani (LB) Lennox broth (pH 6.6) (Sigma-Aldrich, St. Louis, MO) as the  growth medium. The broth consists of tryptone (10 g/L), yeast extract (5 g/L) and NaCl (5 g/L). The ionic strength is $\sim 85~\rm mM$. We also use phosphate buffer saline (PBS) (pH 7.4) (Lonza Walkersville Inc., Walkersville, MD) in some experiments. Details of the procedure for culturing the bacteria can be found in \cite{yang2020all}. We measure the pH values of the media using a Pocket Pro$^{+}$ pH Tester (Hach, Loveland, CO) at $37^{\circ}$C. The incubation of bacteria in the media does not change the pH values, and the pH values of the media remain unchanged over the entire period of our experiment  ($\sim3$ hours).

To fix the cells, the following steps are used.  The cells are first washed twice in PBS,  the washed cells are then fixed in glutaraldehyde (Sigma-Aldrich, St. Louis, MO) in LB for 2 hours at $37^{\circ}$C. The concentration of glutaraldehyde used is 2.5\% (vol/vol) \cite{chao2011optimization}. After this step, the fixed cells are washed twice in PBS and subsequently re-suspended in LB for our noise measurements. Glutaraldehyde is a chemical fixative that crosslinks proteins on bacterial surfaces and inhibits transport processes   \cite{mcdonnell1999antiseptics}. The fixation solution  has been shown to effectively preserve morphology of both bacterial cells and surface ultrastructures for a period longer than the duration of our  experiments ($\sim3$ hours) \cite{chao2011optimization, meade2010studies}. This fixation method is not expected to change the susceptibility of bacteria to electrical perturbations [see Appendix \ref{sec:perturbations}].

\subsection{Experimental setup}

The experiments are performed on an Axio observer inverted microscope (Carl Zeiss, Oberkochen, Germany).  We use a PeCon 2000–2 Temp Controller (PeCon GmbH, Erbach, Germany). At the beginning of each experiment, the sample is loaded into the microfluidic resistor under a pressure-driven flow established through the microfluidic channel using the flow controller OB1-Mk3 (Elveflow, Paris, France). More details regarding sample loading can be found in \cite{yang2020all}.

Our experiments investigate the dependence of the excess noise on three experimental parameters: the amplitude of the bias current $I$, the pressure gradient  $\Delta p$ of the bulk flow, and the number $N$ of bacterial cells. When we focus on one parameter, the other two parameters are maintained more or less unchanged. Briefly, for studying the $I$ dependence, we measure the voltage noise for $0.70~\rm nA \leq$ $I$ $\leq 16.50~\rm nA$, while keeping $N\approx \rm 300 \pm 50$ and $\Delta p \approx 0.6~\rm kPa$. For studying the $\Delta p$ dependence, we measure the voltage noise under constant $\Delta p$ in the range $0.2~\rm kPa \leq$ $\Delta p$ $\leq 3.0~\rm kPa$ (incremented by 0.4 kPa), while we keep $I \approx 7.75~\rm nA$ and $N\approx 300 \pm 50$. For studying the $N$ dependence, we measure the voltage noise with $20 \lesssim N \lesssim 500$, while maintaining $I\approx7.75~\rm nA$ and $\Delta p \approx 0.6~\rm kPa$. All measurements are performed at $37^{\circ}$C, unless indicated otherwise.

\subsection{Electrical measurements and data acquisition}

A SR830 DSP lock-in amplifier  (Stanford Research Systems, Sunnyvale, CA) is used for the measurements. We record the output signals from the lock-in amplifier at a sampling frequency of 128 Hz using a data acquisition card NI 6221 (National Instruments, Austin, TX) through a LabVIEW (National Instruments, Austin, TX) Virtual Instrument interface.  Microscope images of the bacterial cells in the microchannels are acquired on an Axio observer inverted microscope (Carl Zeiss, Oberkochen, Germany) with  5$\times$ and  63$\times$ objectives, an AxioCam 503 mono camera (Carl Zeiss, Oberkochen, Germany), and ZEN image acquisition software (Carl Zeiss, Oberkochen, Germany).  Most data processing is performed using Origin (MicroCal Software, Northampton, MA). 

\section{AUTO-CORRELATION FUNCTION}
\label{app:autocorrelation}

In the  $1/f$ noise literature, the PSD of  noise is   more commonly displayed  than the autocorrelation function---although there are some exceptions \cite{dmitruk2011emergence}. This is partially due to the following reason: to obtain the asymptotic $1/f$ behavior of the noise, one needs to subtract the high-frequency white noise tail, as we have done in Section \ref{sec:basicsteps} (for instance, see Eq. (\ref{eq:normalizedPSD}) and Fig. \ref{fig:figure6}).  Thus, it becomes  harder to visualize asymptotic noise characteristics in the time domain. Regardless, it may be helpful to look at the autocorrelation function here for comparison.

\begin{figure}
    \begin{center}
    \includegraphics[width=3.375in]{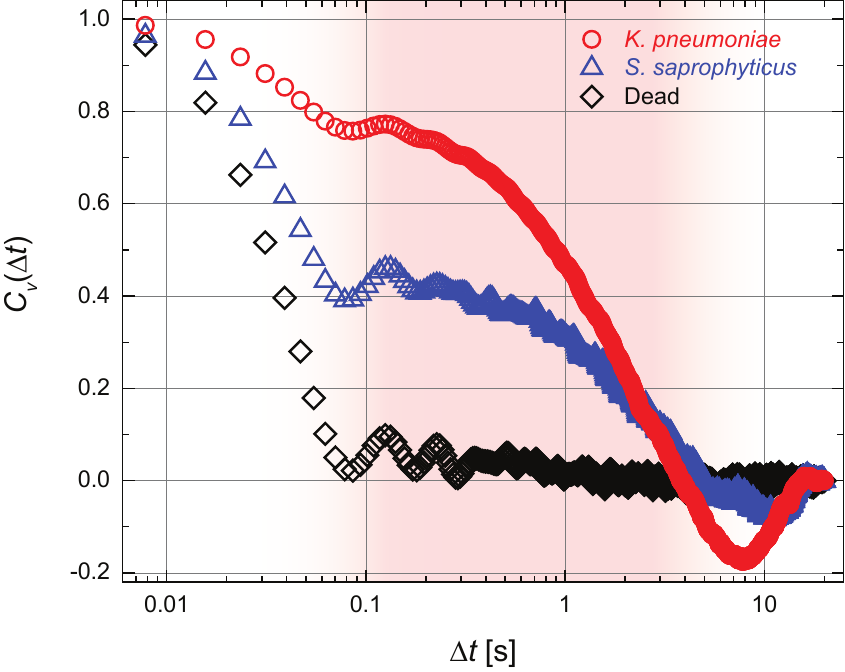} 
    \caption{Normalized temporal autocorrelation functions $C_{v}(\Delta t)$ of voltage fluctuations for live \textit{K. pneumoniae} and \textit{S. saprophyticus} cells and background (dead cells), plotted as a function of  time lag $\Delta t$ in logarithmic scale. The shading approximately indicates the region corresponding to the frequency range shown in the PSDs in Fig. \ref{fig:figure6}.} 
    \label{fig:figAutoCor}
    \end{center}
\end{figure}

Figure \ref{fig:figAutoCor} shows  the normalized autocorrelation functions of the voltage fluctuations,  $C_{v}(\Delta t) = {{{\langle v(t) \cdot v(t+\Delta t) \rangle}} \over {\langle {v(t)}^{2} \rangle}} $, for different conditions. Here, the angular brackets indicate  averaging over time and $\Delta t$ is the time lag. Each data trace in Fig. \ref{fig:figAutoCor} is obtained from 12 20-second-long $v(t)$ data traces. Briefly, $C_{v}(\Delta t)$ of a single trace is computed, smoothed, and averaged.

The increase in $C_{v}(\Delta t)$ at short time scales, $\Delta t < 0.1~\rm s$, is due to the high frequency fluctuations, which are removed from the PSDs. At  large timescales, $\Delta t > 3 s$, $C_{v}(\Delta t)$ becomes inaccurate because  the data traces are only 20 s long. The dip seen in the data, in particular, may be an artifact due to the finite length of the data. The shading in Fig. \ref{fig:figAutoCor} approximately corresponds to the frequency range of the PSDs shown in Fig. \ref{fig:figure6}.     The bacteria data in Fig. \ref{fig:figAutoCor}  can be fit to exponential decays with time constants of $\sim 10 ~\rm s$. In this respect, the autocorrelation function does not provide any new insight into the phenomena.

%*************************************Figure************************
%%%%%%%%%%%%%%%%%%%%%%%%%%%%%%%%%%%%%%%%%%%%%%%%%%%%%%%%%%%%%%%%%%%

\section{THE DIMENSIONLESS FUNCTION $\cal C$}
\label{app:dimensionless}

\begin{figure}
    \begin{center}
     \includegraphics[width=3.375in]{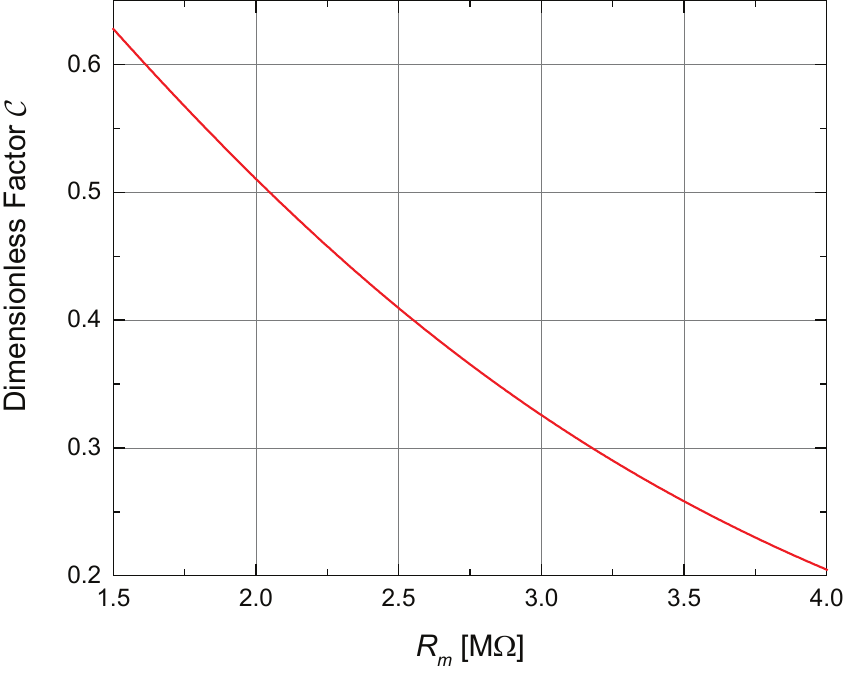} 
    \caption{The dimensionless factor ${\cal C}$  as a function of resistance $R_m$ of the microfluidic resistor.}
     \label{fig:figS1}
     \end{center}
\end{figure}

In Section \ref{sec:excessnoise}, we express the PSD of the current-dependent  excess voltage noise  measured between the nodes $A$ and $B$ in Fig. \ref{fig:figure2}(b) as Eq. (\ref{eq:exPSD}).
%\begin{equation}
%S_V^{(ex)}(f,I)= { \left[ {  \left|{{Z_{in}}} \right|^2 } \over (1+{\omega_o}^2 {R_m}^2 {C_m}^2)^2 {\left| Z_m +Z_{in} \right|}^2 \right] }  I^2 S_R(f) = {\cal C} I^2 S_R(f).   
%\end{equation}
Here, the factor  ${\cal C}$ is a dimensionless coefficient that quantifies how the measured  noise at the input of our electronics is attenuated as compared  to the noise generated in the microfluidic resistor.  In our experiments,  the value of $R_m$  changes with the number $N$ of bacteria in the microchannel and/or the resistivity of the different electrolytes. The parasitic capacitance coming mostly from the wiring stays more or less constant. Thus, the factor ${\cal C}$ is assumed to be only a function of $R_m$. Using the  values given in Section \ref{sec:electricalmeasurments} for $C_m$, $Z_{in}$, and $\omega_o$,  we calculate  ${\cal C}$ as a function of $R_m$, as shown in  Fig. \ref{fig:figS1}. When comparing measurements with different  $R_m$ values, we deconvolute the effects of $R_m$ from  the measured noise for consistency by using the ${\cal C}(R_m)$ corresponding to the $R_m$ value of the microfluidic resistor, as described in Section \ref{sec:electricalmeasurments}.

%  %*************************************Figure************************
% %%%%%%%%%%%%%%%%%%%%%%%%%%%%%%%%%%%%%%%%%%%%%%%%%%%%%%%%%%%%%%%%%%%

\section{PERTURBATIONS, CONTROL EXPERIMENTS, AND VARIOUS OTHER ESTIMATES}
\label{app:perturbations}

\subsection{Electrical perturbations}
\label{sec:perturbations}

At the frequency of the excitation current $f_o=160~\rm Hz$, the expected change in membrane potential $\Delta\Psi$ due to the applied electric field $E(f_o)$ can be calculated using the Schwan equation
\begin{equation}
    \Delta\Psi={{1.5 a E(f_o)}\over{\sqrt{1+(2\pi f_o T)^2}}}.
\end{equation}
Here, $a$ is the radius of the bacterium and $T$ the relaxation time of the membrane \cite{marszalek1990schwan,maswiwat2007simplified}. For a rodlike bacterium like \textit{K. pneumoniae}, we can approximate $a \approx 1~\rm \mu m$ and $2\pi f_o T \ll1$ \cite{xie1990study}. The highest electric field in our experiment, $E(f_o)=350~\rm V/m$, corresponds to fluctuations in membrane potential of $\Delta\Psi_{rms}=0.52~\rm mV$. Thus, electroporation of the bacteria membrane as well as  stimulation of any voltage-gated ion channels are  unlikely \cite{tsong1990electrical,xie1990study,stratford2019electrically}. The largest current density in our experiments ($580~\rm A/m^2$) is also expected to have no harmful effects on the bacteria \cite{sale1967effects,brambach2013current}.

\subsection{Electrokinetic and flow forces on a bacterium}

Here, we compare the magnitude of the various forces acting on the bacteria in the microchannels. In particular, bacteria experience an electrokinetic force induced by the oscillating voltage and a hydrodynamic drag force due to the pressure driven flow. Since these non-motile bacteria are in the Stokes flow regime, estimating the flow velocities will be sufficient to assess the magnitude of the forces. We  thus look at the pressure-driven component, $u_{\Delta p}$, and the electrokinetic component, $u_{EK}$, of the flow velocity in the microchannel in separate control experiments.

\begin{figure*}
    \begin{center}
    \includegraphics[width=6.75 in]{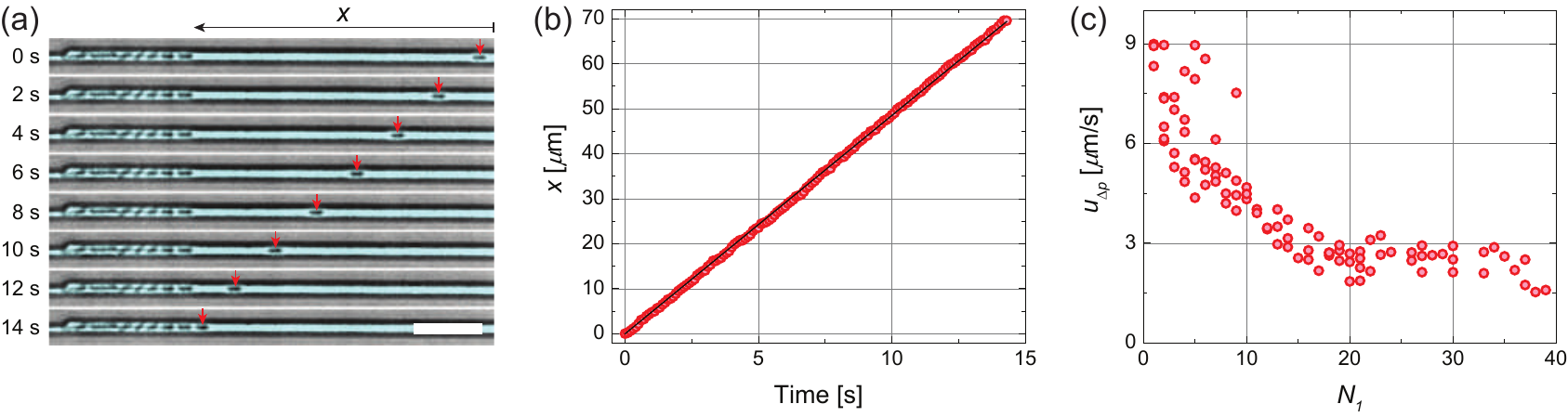}%place holder
    \caption{(a) Time-lapse images of a bacterium in a microchannel in pressure-driven flow. The false colored image shows the broth medium in blue. The scale bar is $10~\rm \mu m$. (b) A set of position \textit{vs.} time data obtained from optical tracking (red symbols). The linear fit (black line) provides the velocity of the bacterium. (c) The flow velocity as a function of the number $N_1$ of bacteria already trapped in a  microchannel. Each data point is determined from a measurement on a single bacterium.}
    \label{fig:figS2}
     \end{center}
 \end{figure*}

The pressure-driven velocity $u_{\Delta p}$ at $\Delta p \approx 0.6~\rm kPa$ can be estimated through optical tracking. During an experiment with live bacteria,  bacteria entering the microchannels  are optically tracked, as shown in Fig. \ref{fig:figS2}(a). We use ImageJ \cite{rasband2011imagej} to analyze the videos which have a frame-rate of $10~\rm fps$ and a resolution of $\pm 200~\rm nm$. Each bacterium is tracked from the moment it enters the microchannel until its collision with the  bacteria already trapped by the nanoconstriction [Fig. \ref{fig:figS2}(a)]. The velocity is obtained by a linear fit of the position over time  as shown in Fig. \ref{fig:figS2}(b). The velocity depends on the number $N_1$ of bacteria  already trapped in the  microchannel [Fig. \ref{fig:figS2}(c)], because each bacterium trapped in the microchannel increases the hydraulic resistance of the microchannel. A microchannel gets completely filled with bacteria when $N_1\approx40$. During the experiments, about $300-350$ bacteria are trapped in all ten microchannels. From Fig. \ref{fig:figS2}(c), we estimate the steady-state flow velocity during electrical noise measurements to be $u_{\Delta p}\approx2.5~\rm \mu m/s$.

% %*************************************Figure************************
% %%%%%%%%%%%%%%%%%%%%%%%%%%%%%%%%%%%%%%%%%%%%%%%%%%%%%%%%%%%%%%%%%%%

\begin{figure*}
    \begin{center}
     \includegraphics[width=6.75in]{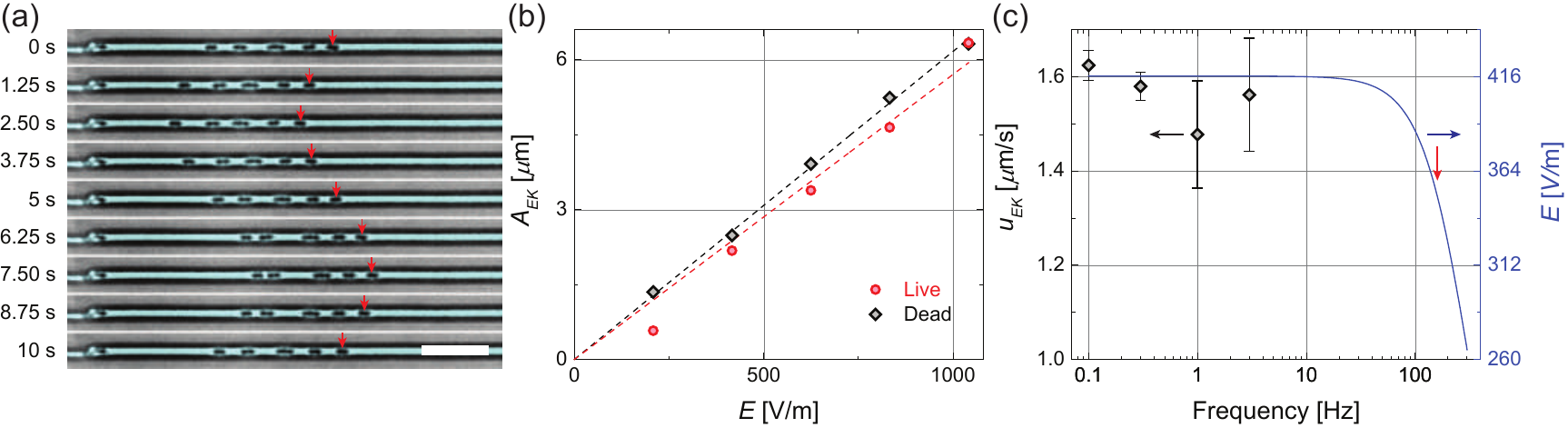}%place holder
     \caption{(a) Time-lapse images of the electrokinetic movements of bacteria in a microchannel. Here, the velocity $u_{EK}$ is determined from  optical tracking of these movements at $f=0.1~\rm Hz$ and $E=1,041~\rm V/m$. The false colored image shows the broth medium in blue. The scale bar is $10~\rm \mu m$. (b) Comparison showing that electrokinetic forces acting on live and dead bacteria are very close in magnitude. Here, we measure the displacement amplitude $A_{EK}$ of live and dead bacteria at different electric field strengths at a frequency of $0.1~\rm Hz$. (c) Electrokinetic velocity as a function of frequency. The black symbols show measurements taken at four different frequencies. The blue curve shows the frequency dependence of $E(f)$. The red arrow  indicates that the largest electric field strength used in our experiments at $f_o=160~\rm Hz$  is  $E(f_o)=350~\rm V/m$. }
     \label{fig:figS3}
     \end{center}
 \end{figure*}

The electrokinetic flow velocity is measured in separate experiments by tracking the sinusoidal displacements of single live and dead bacteria at various carrier frequencies $f$ and electric field strengths $E$ in the absence of any pressure-driven flow. Figure \ref{fig:figS3}(a) shows an example of the oscillation cycle of a dead bacterium at $f=0.1~\rm Hz$ with $E=1,041~\rm V/m$. Bacteria are observed for $30~\rm s$ with a framerate of $30~\rm fps$, and the displacement is again tracked using ImageJ. Figure \ref{fig:figS3}(b) shows the displacement amplitudes $A_{EK}$ of a live and a dead bacterium collected at $f=0.1~\rm Hz$ for five different electric field strengths. The dashed lines show a linear fit passing through the origin. No significant changes in $A_{EK}$ are found between live and dead bacteria, suggesting that neither electrokinetic forces nor friction coefficients are affected by fixation with glutaraldehyde. 

Our goal is to determine the largest value of the electrokinetic velocity in our experiments obtained at an applied electric field strength of $E(f_o)=350~\rm V/m$ at the carrier frequency of $f_o=160~\rm Hz$. Because of resolution limits, we cannot directly measure this quantity. Instead, we  measure the displacement amplitudes of a single bacterium at low frequencies and high electric fields. The value of $E(f_o)$ is slightly lower than $E(f\lesssim10~\rm Hz)$ due to attenuation by the parasitic capacitance in our system, as shown by the calculated blue curve in Fig. \ref{fig:figS3}(c). For $f\leq3~\rm Hz$, we convert the displacement amplitudes to velocity via $u_{EK}(f)=2\pi f A_{EK}(f)$. We find $u_{EK}(f\leq3~\rm Hz) \approx 1.6~\rm \mu m/s$, corresponding to an electrokinetic mobility of $\mu_{EK}\approx3.8\times10^{-9}~\rm m^2/(V\cdot s)$ through the relation $u_{EK}(f)\approx\mu_{EK}E(f)$.  Using this mobility value, we find the maximum electrokinetic velocity at 160 Hz to be $u_{EK}(f_o)\approx1.3~\rm \mu m/s$.

% %*************************************Figure************************
% %%%%%%%%%%%%%%%%%%%%%%%%%%%%%%%%%%%%%%%%%%%%%%%%%%%%%%%%%%%%%%%%%%%

In summary, the electrokinetic force acting on the bacteria is measured to be smaller than the drag force due to the pressure-driven flow. Hence, the bacteria are pushed toward the nanoconstrictions in the microchannels. Once loaded under the high pressure flow, the cells tend to stay in place and not move at all.

\subsection{Time constants}
\label{app:time}

There are a number of time constants in the system, in addition to the charge relaxation time of a single bacterium. Here, we discuss these and provide estimates for each. Since there is a steady flow of velocity $2.5~\rm \mu m$ from inlet to outlet, the microchannels get flushed in a time scale  $\tau_f \sim {l\over {u_{\Delta p}} } \sim 40~\rm s$. The diffusion time constant, $\tau_d \sim {l^2\over D} \sim 10 ~\rm s$, where $D\sim 10^{-9}~\rm m^2/s$ is the diffusion coefficient for small inorganic  cations in water \cite{samson2003calculation}. The charge relaxation time of the bacteria and $\tau_d$ are of the same order. Finally,  ions drift in the microchannels due to the electric field. Since we are using an ac field,  the cations drift on the average $l_\mu \sim {\mu E(f_o) \over 2f_o} \sim 110 ~\rm nm$ during  half of a cycle of the oscillating field. Here, the mobility for small cations is taken as $\mu \sim 10^{-7}~\rm m^2/(V \cdot s)$ \cite{kirby2010micro} and the largest electric field value is used for the estimate.   Since $\l_{\mu} \ll l$, drift does not play a role in our observations. 

\subsection{Temperature increase due to Joule heating}

We estimate the temperature increase in our microchannels due to Joule heating using a control volume approach. We use one of the ten channels as the control volume. Conservation of energy per unit time in steady state leads to:
\begin{equation}
    \dot{m}c_p(T_\infty-T_{C})+{I_1}^2R_1-q''A=0.
\end{equation}
Here, $\dot{m}$ is the mass flow rate, $c_p$ is the specific heat capacity of the broth, $T_\infty=310~\rm K$ is the temperature of our sample far removed from the channel, $T_{C}$ is the microchannel temperature, $I_1={{I}\over{10}}\approx1.7~\rm nA$ is the maximum current passing through the channel, $R_1=10R\approx31~\rm M\Omega$ is the channel resistance, $q''$ is the heat flux, and $A$ is the relevant area of channel wall. The mass flow rate can be expressed as $\dot{m}=\rho_lwhu_{\Delta p}$, where $\rho_l\approx1,000~\rm kg/m^3$ is the density of the broth. We use $c_p\approx4,000~\rm J/(kg\cdot K)$ \cite{han2004thermodynamic}. We simplify the calculation of $q''$ using a one-dimensional model by neglecting any heat flux through the PDMS due to the smaller thickness $d=1~\rm mm$ and higher thermal conductivity  of the glass substrate \cite{erickson2003joule}. The heat flux through the glass substrate can then be approximated as
\begin{equation}
    q'' \approx k \left ({{T_C-T_\infty}\over{d}} \right ).
\end{equation}
Here, $k=1.4~\rm W/(m\cdot K)$ is the thermal conductivity of glass. With the contact area between broth and glass substrate $A=l w$, the channel temperature can then be expressed as
\begin{equation}
    T_C=T_\infty+{{{I_1}^2R_1}\over{\dot{m}c_p+k{{lw}\over{d}}}}.
\end{equation}
For the maximum current used in our experiments, the temperature in the microchannels increases by $\sim3.2\times10^{-4}~\rm K$.

\subsection{Resistance fluctuations due to  nanomechanical fluctuations of a bacterium}

\subsubsection{Finite element model}

\begin{figure*}%[!h]
     \begin{center}
     \includegraphics[width=5.55in]{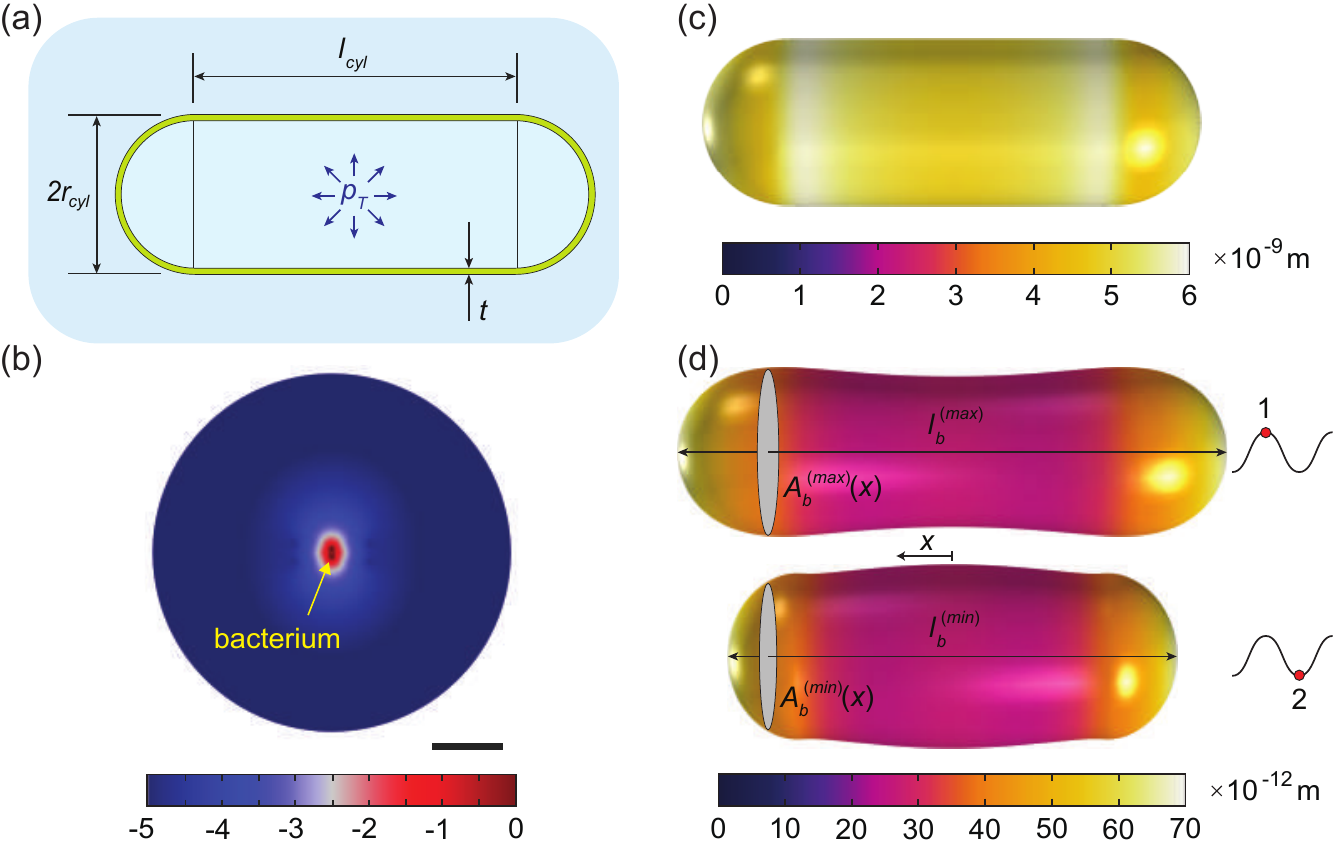}%place holder
     \caption{(a) Cross-section of a bacterium (\textit{K. pneumoniae}) immersed in water. The bacterium is modeled as a hollow cylinder with wall thickness $t$, length $l_{cyl}$ and radius $r_{cyl}$, with a hollow semi-sphere of the same thickness on each end. The inside of the bacterium is modeled as water under turgor pressure $p_T$ \cite{zhang2021measurement}. (b) In the model,  the bacterium is immersed in a sphere of water with $100~\rm \mu m$ diameter to capture all viscous, acoustic and thermal losses. The logarithmic plot of the normalized fluid velocity  shows that the fluid can be assumed as quiescent at the boundary of our model. The scale bar is $20~\rm \mu m$. (c) Steady-state expansion due to turgor pressure. The induced tension in the cell wall affects the eigenmode oscillations. (d) Maximum (1) and minimum (2) deformations during the oscillation cycle of the eigenmode at $8.26~\rm MHz$. The amplitude field shown gives a  strain energy  that is equal to the thermal energy at $310~\rm K$. The respective lengths and cross-sections of the bacterium are represented as $l_b^{(max)}$, $l_b^{(min)}$, $A_b^{(max)}$, and $A_b^{(min)}$. The eigenmode deformations are artificially scaled up by a scale factor of $2000$ for enhancing the contrast.}
     \label{fig:figS4}
     \end{center}
 \end{figure*}

A finite element model of the nanomechanical dynamics of a \textit{K. pneumoniae} cell in water is created using COMSOL Multiphysics\texttrademark. The bacterium is modeled as a hollow cylinder with length $l_{cyl}=2~\rm \mu m$, radius $r_{cyl}=500~\rm nm$ and wall thickness $t=26~\rm nm$, with two semi-spheres of the same radius and thickness attached to both ends, as shown in Fig. \ref{fig:figS4}(a) \cite{zhang2021measurement}. The cell wall is modeled as a linear elastic material with Young's modulus $E=49~\rm MPa$, Poisson's ratio $\nu=0.16$, and density $\rho_b=1,100~\rm kg/m^3$ \cite{mirzaali2018silico,zhang2021measurement,gerhardt1964porosity}. The inside of the bacterium is modeled as water under turgor pressure ($p_T=29~\rm kPa$) \cite{zhang2021measurement}. The bacterium is surrounded by a sphere of water under atmospheric pressure. Both bodies of water are modeled as viscous compressible Newtonian fluids, and the diameter of the surrounding water sphere  is chosen as $100~\rm \mu m$ such that all viscous and thermal losses can be captured within the model [Fig. \ref{fig:figS4}(b)] \cite{liem2021nanoflows}. A pre-stressed eigenfrequency study is used, which consists of a stationary solver followed by an eigenfrequency solver. For the stationary step, a boundary load is placed on the inner wall of the bacterium to model the expansion due to the turgor pressure. For the eigenfrequency step, a multiphysics coupler connects the Solid Mechanics and Thermoviscous Acoustics modules. 

%*************************************Figure************************
%%%%%%%%%%%%%%%%%%%%%%%%%%%%%%%%%%%%%%%%%%%%%%%%%%%%%%%%%%%%%%%%%%%

The matrix ${\bf{x}}_0$ contains the initial radial and axial coordinates for each of the $N$ finite elements comprising the axisymmetric cell surface. The turgor pressure expands the shell coordinates to ${\bf{x}}_0^{'}$ [Fig. \ref{fig:figS4}(c)]. The time dependent coordinates of the $N$ finite elements during eigenmode oscillations are of the form  ${\bf{x}}_0^{'}+{\bf{a}}_n \cos{\omega_n t}$, where ${\bf{a}}_n$ is the modal amplitude, i.e., the amplitude along the eigenvector, and $\omega_n$ is the eigenfrequency. An example of the deformations in one of the eigenmodes is shown in Fig. \ref{fig:figS4}(d).

In order to make more quantitative comparisons, we will use the strain energy $U$.  Strain energy will allow us to express a given rms oscillation amplitude in units of the strain energy $U^{(th)}$  of the thermal fluctuations and define an approximate effective temperature.  To this end, we can easily find  the modal amplitude ${\bf{a}}_n ^{(th)}$ for each mode, which results in a strain energy of  $U^{(th)} = k_BT$, where $k_B$ is the Boltzmann constant and $T=310~\rm K$ is the equilibrium temperature. One could think of $U^{(th)}$ as the approximate energy of a fixed  bacterium---assuming the physical properties of the fixed bacterium do not change upon chemical fixing. 

In live bacteria, nanomechanical fluctuations originating from active processes must be dominant compared to the thermal fluctuations. Nanoscale movements due to active processes has been observed in  different  human and animal cells \cite{ben2011effective,turlier2016equilibrium,biswas2017mapping,nelson2017vibrational,wu2016quantification,singh2019studying},  giant vesicles \cite{manneville2001active}, yeast \cite{willaert2020single}, as well as motile and non-motile bacteria \cite{roslon2022probing,longo2013rapid}. For non-motile bacteria, these active processes include, but are not limited to, the activity of ion pumps \cite{dimroth2006catalytic}, the movement of proteins across the outer membrane \cite{lenn2008clustering,spector2010mobility}, and  the fluctuations in the cytoskeleton (e.g., due to motor proteins) \cite{winkler2020physics}.  A live bacterium would thus have a strain energy $U^{(act)}$ that is significantly larger than the thermal strain energy  $U^{(th)} = k_BT$, discussed above. The corresponding effective temperature,  $k_B T_{eff}= U^{(act)}$, is also larger than the equilibrium temperature $T=310~\rm K$.  

\subsubsection{Conversion of deformations into resistance changes}

The rms resistance change $r$ across a single microchannel with cross-sectional area $A_c=4~\rm \mu m^2$ and length $l_c=100~\rm \mu m$ caused by one bacterium oscillating coherently with an energy equal to  its equilibrium thermal energy can be approximated as 
%begin{equation} 
%\begin{split}
%    r & \approx {{\rho}\over{2\sqrt{2}}} \left| {\int\limits_{-l_b^{(max)}/2}^{l_b^{(max)}/2} {{{dx} \over {{A_c} - {A_b^{(max)}}(x)}}} } \right. \\
%   & \left. - { {\int\limits_{-l_b^{(min)}/2}^{l_b^{(min)}/2}} {{{dx} \over {{A_c} 
%    - {A_b^{(min)}(x)}}}}} - {{l_b^{(max)}-l_b^{(min)}}\over{A_c}}\right|.
%    \label{eq:r}
%    \end{split}
%\end{equation}
\begin{widetext}
\begin{equation} 
    r \approx {{\rho}\over{2\sqrt{2}}} \left| {\int\limits_{-l_b^{(max)}/2}^{l_b^{(max)}/2} {{{dx} \over {{A_c} - {A_b^{(max)}}(x)}}}  - {\int\limits_{-l_b^{(min)}/2}^{l_b^{(min)}/2}} {{{dx} \over {{A_c} - {A_b^{(min)}(x)}}}}}  - {{l_b^{(max)}-l_b^{(min)}}\over{A_c}}\right|.
    \label{eq:r}
\end{equation}
\end{widetext}
Here, $r$  is calculated by finding the difference in microchannel resistance between the maximum and minimum deformation states of the bacterium [shown as 1 and 2 in Fig. \ref{fig:figS4}(d)]. The parameters $l_b^{(max)}$, $l_b^{(min)}$, $A_b^{(max)}$, and $A_b^{(min)}$ can be understood as the respective bacterial lengths and cross-sections of these deformed states. The value of $\rho$ for LB broth has been determined as $\rho\approx 1.2~\rm \Omega \cdot m$   from previous measurements \cite{yang2020all}. To account for the $300 \pm 50$ bacteria in our experiments, we assume 30 bacteria  in each of the ten microchannels, all oscillating in the same eigenmode.  In this step, by assuming that the oscillations are not coherent but their noise powers are additive, we  make a transition from  eigenmode oscillations to fluctuations. We calculate the total rms resistance fluctuations in our system as $r_{300}\approx {{\sqrt{30}}\over{10}}r$.

We show the obtained values for $r_{300}$ for four different eigenmodes in Table \ref{tab_simulation}. Here, the modal strain energies are set to the thermal energy $U^{(th)}$ as described above, and hence the effective temperature is $310~\rm K$. In order to compare the simulated results with the resistance fluctuations observed in our measurements, we  find the experimental rms resistance fluctuations by calculating the variance of the resistance fluctuations from the measured PSDs [see Fig. \ref{fig:figure3}] in the frequency range ${0.05~\rm Hz} \leq f \leq {1~\rm Hz}$ and properly subtracting the background noise power. The experimentally-determined value is $r_{300} \approx 214~\rm \Omega$. We then find the  modal amplitude ${\bf{a}}_n ^{(hot)}$ from numerical simulations, at which the simulated $r_{300}$ matches the experimental value based on Eq. (\ref{eq:r}) and $r_{300}\approx {{\sqrt{30}}\over{10}}r$. We do this exercise for the eigenmode with the largest thermal $r_{300}$ value, i.e., the eigenmode at $8.26~\rm MHz$ with $r_{300}\approx 0.766~\rm \Omega$.  The effective temperature $T_{eff}$ at which active fluctuations attain  the experimentally measured $r_{300}$ is determined by using the strain energy ratios as ${{{U^{(act)}}} \over {{U^{(th)}}}} = {{{T_{eff}}} \over {310}}$.  As a result,  we obtain that $T_{eff}\sim 10^7~\rm K$ for the bacteria in the experiments.

\subsubsection{Discussion}

Our eigenfrequency and quality factor  values for the first mode generally agree  with those in other simulations where bacteria are modeled as floating shells filled with and surrounded by water \cite{zinin2009deformation,choi2010comment,tamadapu2015resonances}; they  are, however, significantly lower than the values obtained in simulations where bacteria are modeled as  solid spheres attached to a substrate and surrounded by air  \cite{gil2020optomechanical}. We also note that the following factors  seem to affect the results in the literature: (i)~the material properties of the bacterial shell; (ii)~ boundary conditions; (iii)~computing algorithm. We are  not sure which physical mechanisms dominate the energy dissipation. We also don't know whether or not the $Q$ factors observed here and in the literature are reasonable.

Because of a lack of understanding of the nanomechanical fluctuations of bacteria at low frequencies, our model comes with shortcomings and only provides some level of comparison. \textit{First}, we are essentially comparing thermal fluctuations with $1/f$ noise. Thermal fluctuations in this system are  spread over a broad bandwidth on the order of a few MHz, as suggested by the $Q$ values, whereas the observed $1/f$ noise is at low frequency within 2 Hz from  dc. \textit{Second}, we assume that the active  cell movements come with the spatial deformations  of the eigenmodes---even though active processes should excite the cell deformations at random locations. Regardless, the enormous difference between the observed and computed strain energies has convinced us that nanomechanical motion is an unlikely source for our observations.

 \begin{table*}
 \caption{\label{tab_simulation} Estimated parameters for the simulated nanomechanical eigenmodes of \textit{K. pneumoniae}. We only show selected eigenmodes which have the largest deformations. The first column shows the deformation of the bacterium in the eigenmode. The deformations are artificially scaled up by a scale factor of $2000$ for better contrast. The second and third columns show the eigenfrequency and the quality factor of the mode, respectively. The fourth column  lists the effective temperature for the strain energy, determined by equating the strain energy to $k_B T_{eff}$, as described in the text. The last column is the rms value of the resistance fluctuations $r_{300}$ caused by 300 bacteria at a strain energy of $k_B T_{eff}$.  The bottom row shows the experimental values for comparison.}
 \begin{ruledtabular}
 \begin{tabular}{cccccc}

 & Eigenmode & $f$ & $Q$ & $T_{eff}$ & $r_{300}$\\
 & & $[\rm MHz]$ &  & $[\rm K]$ & $[\Omega]$\\

 \hline\\
 \multirow{-2}{-30pt}{\includegraphics[height=150 pt]{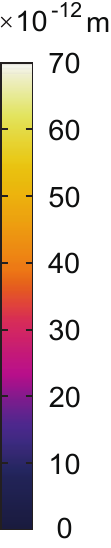}}
 &\raisebox{-.4\totalheight}{\includegraphics[width=0.155\textwidth]{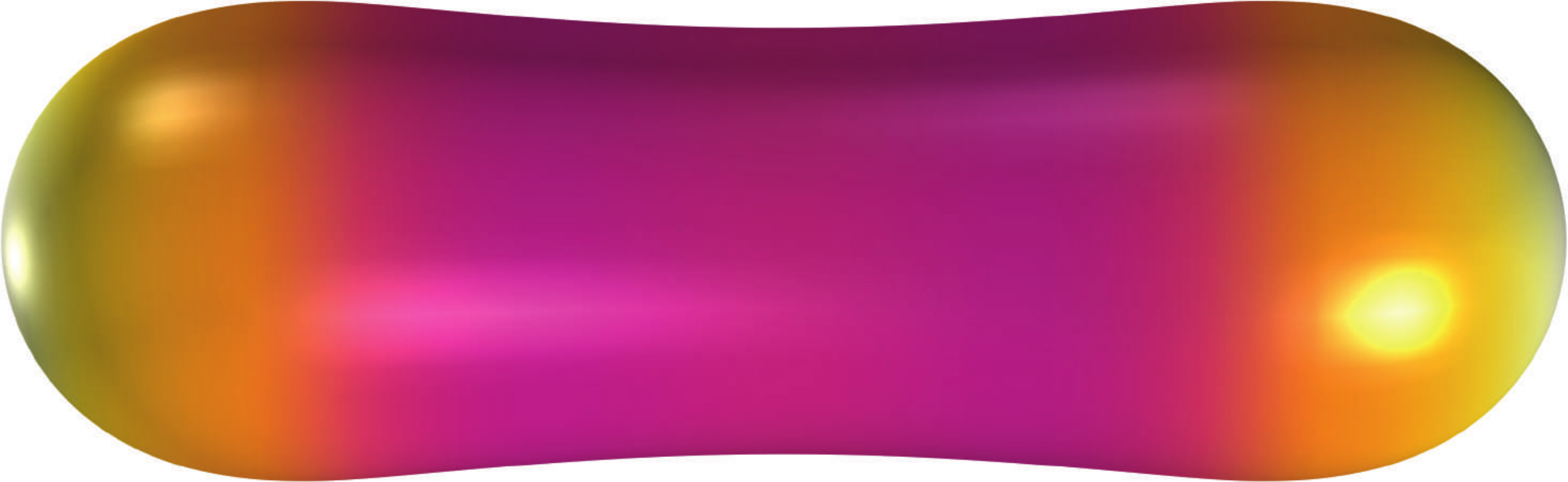}} & $8.26$ & $3.09$ & $310$ & $0.766$\\
 \\
 & \raisebox{-.4\totalheight}{\includegraphics[width=0.155\textwidth]{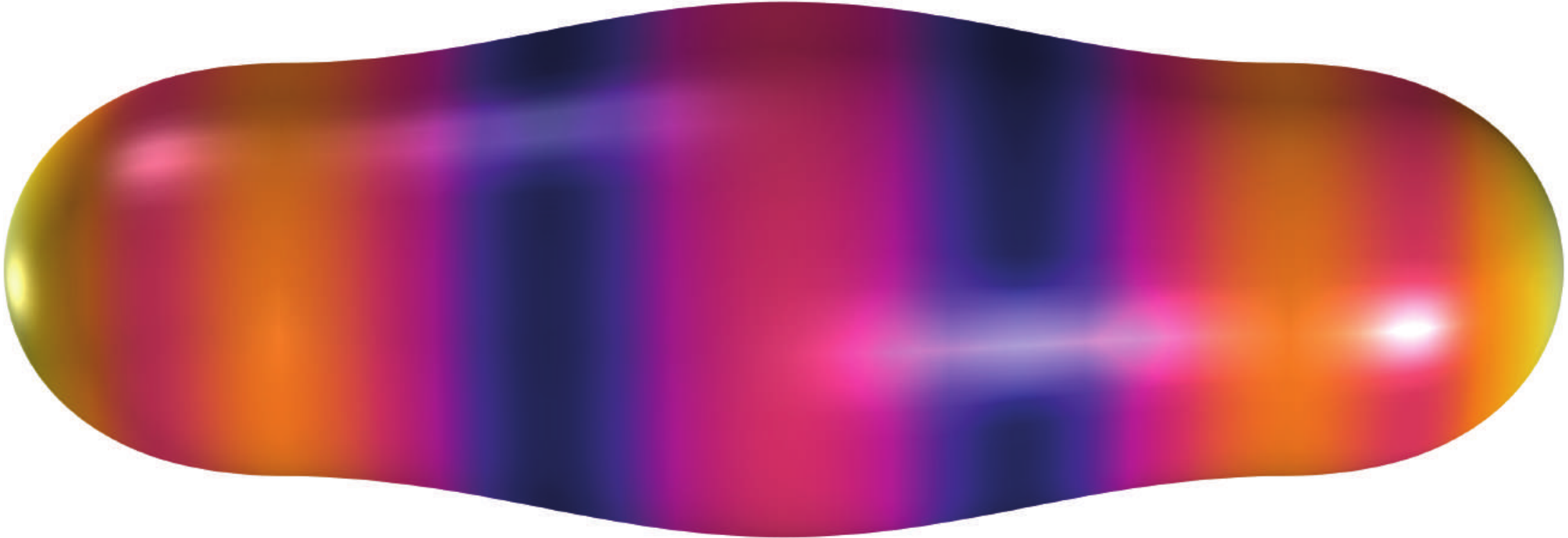}} & $11.88$ & $2.19$ & $310$ & $0.524$\\
 \\
 & \raisebox{-.4\totalheight}{\includegraphics[width=0.155\textwidth]{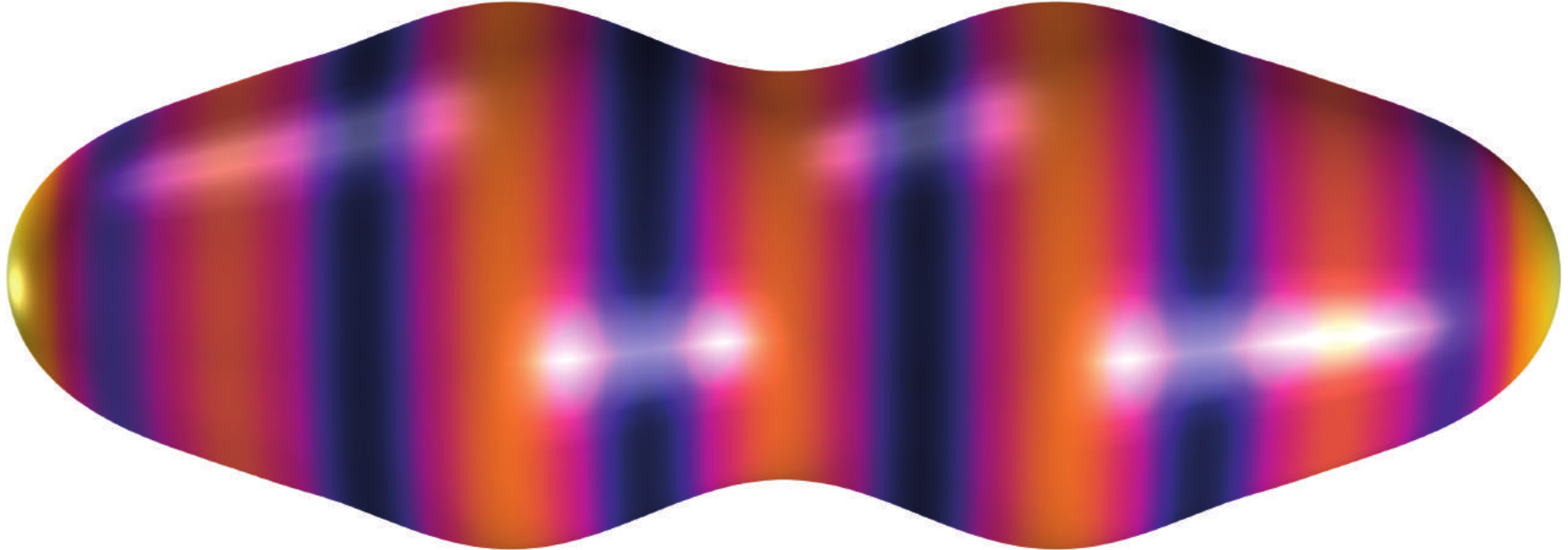}} & $15.95$ & $2.15$ & $310$ & $0.219$\\
 \\
 & \raisebox{-.4\totalheight}{\includegraphics[width=0.155\textwidth]{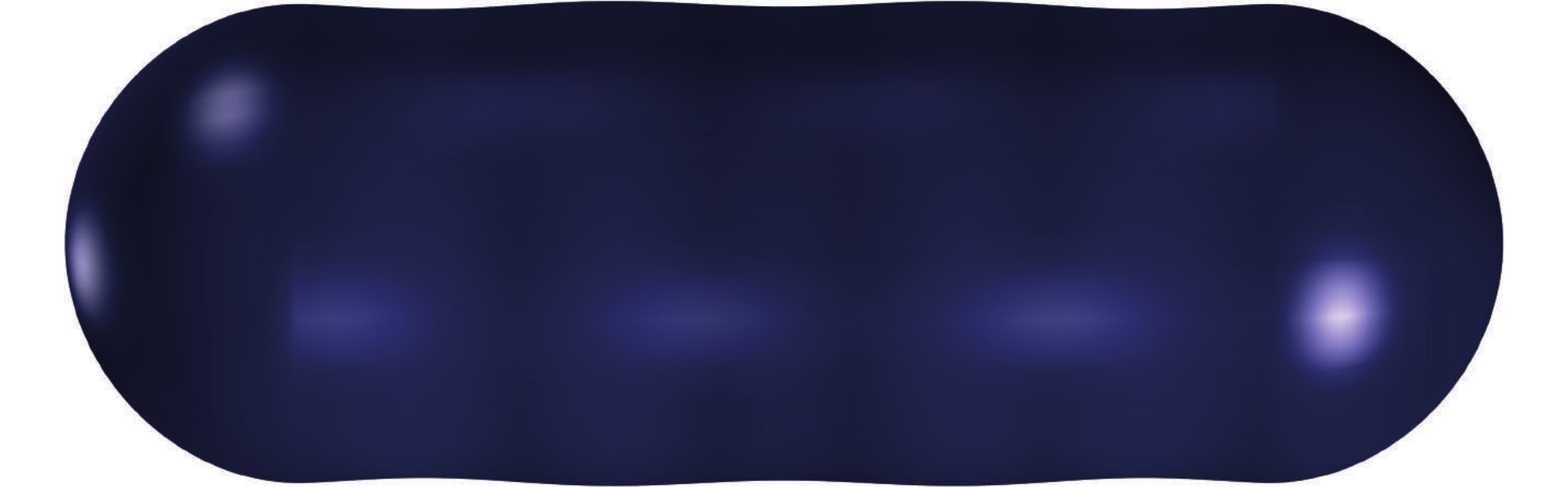}} & $18.48$ & $1.69$ & $310$ & $0.012$\\
 \\
 \hline\\
 & Experiment & - & - & $2.4\times10^7$ & 214.0

 \label{tab:tabS2}
 \end{tabular}
 \end{ruledtabular}
 \end{table*}

\section{ELECTRICAL NOISE CALCULATIONS}
\label{app:chargenoise}

\subsection{Determining charge noise from resistance noise}
As noted in the main text, the relationship between the PSD of the resistance fluctuations, $S_R (f)$, and the PSD of the carrier number fluctuations, $S_n (f)$, is given by Eq. (\ref{eq:eqpsd}).
%\begin{equation}
%    S(f) = {{{S_R}(f)} \over {{{R_m}^2}}} = {{{S_n}(f)} \over {{n}^2}}.
%\end{equation}
As shown in Eq. (\ref{eq:eqpsd}), both $S_R (f)$ and $R_m$ are resultant values for the entire microfluidic resistor, i.e., the ten microchannels in parallel. It is easy to show that ${{{S_R}(f)} \over {{{R_m}^2}}} = {{{S_R^{(1)}}(f)} \over {{{R_1}^2}}}$, where  $R_1$ and ${S_R^{(1)}}(f)$   are respectively the resistance value and the PSD of the resistance fluctuations  of a single microchannel (out of the ten). Since $R_1=\rho {l \over A_c}=\rho {l \over wh}$, $R_m= \rho {l \over 10 A_c}$, ${S_R^{(1)}}(f)={S_\rho}{l^2 \over {A_c}^2}$ and ${S_R}(f)={S_\rho}{l^2 \over 100 {A_c}^2}$, we arrive at $S(f)={{{S_R}(f)} \over {{{R_m}^2}}} = {{{S_R^{(1)}}(f)} \over {{{R_1}^2}}}={S_\rho \over \rho^2}$  for spatially uniform fluctuations. Here, $S_\rho$ is  the PSD of the resistivity fluctuations.  We  make estimates by focusing  on  a single microchannel (out of the ten), in which  $n\approx 4 \times 10^{10}$ at  the $85$ mM  NaCl concentration of LB. There are $N\approx 30$ bacteria in a single microchannel. The result holds if we focus on the entire microfluidic resistor, where $n\approx 4 \times 10^{11}$ and $N\approx300$.

\subsection{Estimating voltage noise from charge noise}
\label{app:voltagenoise}

To find the value of $\tau$, we turn to previous work \cite{kralj2011electrical}. In Fig. \ref{fig:figure3} of \cite{kralj2011electrical}, the authors have shown that  spontaneous electrical blinks in bacteria decay exponentially on  timescales ${10~\rm s} \lesssim \tau \lesssim {30~\rm s}$. These timescales are obtained by fitting the autocorrelation function of the fluorescence intensity measured from single bacterial cells to a single exponential decay. For $\tau ={10~\rm s}$ and ${30~\rm s}$, we find the rms value of the fluctuations in the number of ions for one bacterium is ${\Delta n^{(1)}_{rms}} \approx 1.3 \times 10^{6}$, based on our data shown in Fig. \ref{fig:figure6} in the main text.

In the second approach provided in the main text, we estimate the noise, $e_n$, in the membrane potential, $V_{mem}$, from the fluctuations in the intracellular ion concentrations. We assume that the ions are distributed uniformly inside and outside the cell. We focus on the $\rm K^+$, $\rm Na^+$, and $\rm Cl^-$  because these three  make the largest contribution to the steady-state value of $V_{mem}$ (i.e., the resting membrane potential). The Goldman-Hodgkin-Katz  (GHK) equation \cite{benarroch2020microbiologist} provides the value of $V_{mem}$ as
\begin{widetext}
\begin{equation}
    {V_{mem}} =  {{RT} \over F } {\ln{\left({{p_{\rm K}[{\rm K^+}]_o}+{p_{\rm Na}[{\rm Na^+}]_o}+{p_{\rm Cl}[{\rm Cl^-}]_i}}\over {{p_{\rm K}[{\rm K^+}]_i}+{p_{\rm Na}[{\rm Na^+}]_i}+{p_{\rm Cl}[{\rm Cl^-}]_o}} \right)}}.
    \label{eq:GHK}
\end{equation}
\end{widetext}
Here, $R=8.314~\rm J\cdot K^{-1} \cdot mol^{-1}$ is the universal gas constant, $T\approx 310~\rm K$ is the temperature, $F=96,485~\rm C \cdot mol^{-1}$ is  Faraday's constant; $p_{K}$, $p_{Na}$, and $p_{Cl}$ are respectively the relative membrane permeabilities; $[{\rm K^+}]_i$,  $[{\rm Na^+}]_i$, and $[{\rm Cl^-}]_i$ are respectively the  intracellular ion concentrations; and $[{\rm K^+}]_o$, $[{\rm Na^+}]_o$, and $[{\rm Cl^-}]_o$ are respectively the uniform extracellular ion concentrations for  $\rm{K^+}$, $\rm{Na^+}$, and $\rm{Cl^-}$.

For a bacterium, the relative membrane permeabilities are ${p_{\rm K}}:{p_{\rm Na}:{p_{\rm Cl}}}=1:0.05:0.45$; the  intracellular ion concentrations are $[{\rm K^+}]_i=150 ~\rm mM$, $[{\rm Na^+}]_i=15 ~\rm mM$, and $[{\rm Cl^-}]_i=10 ~\rm mM$; the  extracellular ion concentrations are $[{\rm K^+}]_o=4 ~\rm mM$, $[{\rm Na^+}]_o=145 ~\rm mM$, and $[{\rm Cl^-}]_o=110 ~\rm mM$. By substituting the values of the relative permeabilities  $p_{\rm X}$, the  intracellular ion concentrations  $[{\rm X}]_i$, and the extracellular ion concentrations  $[{\rm X}]_o$ into Eq. (\ref{eq:GHK}), we find  the steady-state value of the membrane potential to be $V_{mem} \approx -67.92~\rm mV$. For a given ion $\rm X$ with charge $z$, we use $E_{\rm X}= {{RT} \over zF } \ln \left([{\rm X}]_o \over [{\rm X}]_i \right)$ to find the equilibrium potentials (Nernst potentials).  This yields  $-96.81~\rm mV$, $60.60~\rm mV$, and $-64.05~\rm mV$  for $\rm{K^+}$, $\rm{Na^+}$, and $\rm{Cl^-}$, respectively. 

To find the total rms change in $V_{mem}$ due to fluctuations in  the concentration of each ion, we calculate the change (fluctuation) in potential from Eq. (\ref{eq:GHK}) with respect to each $[{\rm X}]_i$ as ${{\partial V_{mem}} \over {\partial [{\rm X}]_i}} \Delta [{\rm X}]_i $, square each fluctuation value, add the squares, and then take the  square root of the sum. Then, we find the total rms change $e_n$ described by Eq. (\ref{eq:eqen}).
%\begin{equation}
%    {e_n} \approx  {{RT} \over F } {\left({{{\left(p_{\rm K} \Delta [{\rm K^+}]_i\right)^2} +{\left(p_{\rm Na} \Delta [{\rm Na^+}]_i\right)^2} }\over {\left({p_{\rm K}[{\rm K^+}]_i} +{p_{\rm Na}[{\rm Na^+}]_i}+{p_{\rm Cl}[{\rm Cl^-}]_o}\right)^2}} + {{\left(p_{\rm Cl} \Delta [{\rm Cl^-}]_i\right)^2} \over {\left({p_{\rm K}[{\rm K^+}]_o} +{p_{\rm Na}[{\rm Na^+}]_o}+{p_{\rm Cl}[{\rm Cl^-}]_i}\right)^2}} \right)^{1/2}}.
%   \label{eq:GHKnoise}
%\end{equation}
As shown in Eq. (\ref{eq:eqen}), $\Delta [{\rm K^+}]_i$, $\Delta [{\rm Na^+}]_i$, and $\Delta [{\rm Cl^-}]_i$ are the rms fluctuations in the intracellular ion concentrations; the other parameters are as listed above for Eq. (\ref{eq:GHK}). For a bacterium with  volume $V_b \sim 10^{–18}~\rm m^{-3}$, the rms fluctuations in intracellular ion concentration for each ion is estimated from our experiments to be of order $\Delta [{\rm K^+}]_i=\Delta [{\rm Na^+}]_i=\Delta [{\rm Cl^-}]_i \sim {{{\Delta n^{(1)}_{rms}}} \over {V_b} } \sim 10^{24} ~\rm m^{-3}$. The extracellular ion concentrations are assumed to remain unchanged. Then, by substituting all the values into Eq. (\ref{eq:eqen}), we find $e_n \sim 1.3~{\rm mV}$.

\section{SYMBOLS USED}
\label{app:symbols}

Below we list all and define all the symbols used throughout the main text and the Appendices.

\begin{center}

%\begin{ruledtabular}

\begin{longtable*}{c|lcccccccc}
%\begin{longtable}[c]{| c | c |}

%\caption{Long table caption.\label{long}}\\
%\hline

Symbol & Definition    \\ \hline\\

$ a$ &   Radius of a bacterium  \\
$ {\bf{a}}_n$ &   Modal amplitude of  a bacterium \\
$ A $ &   Relevant area of microchannel wall  \\
$A_b$ & Cross-sectional area of a bacterium \\
$A_c$ & Cross-sectional area of a microchannel \\
$ A_{EK}$ &   Displacement amplitude of bacteria due to the applied electric field \\
$ c_p$ &    Specific heat capacity of the broth \\
$ {\cal C}$  & Dimensionless correction factor     \\
$ C$  &  Membrane capacitance of the cell   \\
$ C_{in}$  &   Equivalent capacitance of the lock-in amplifier  \\
$C_m$  & Equivalent capacitance of the microﬂuidic resistor    \\
$ d$ &    Thickness of the glass substrate\\
$ D$ &   Diffusion coefficient for small inorganic  cations in water  \\
$e$  & Elementary charge    \\
${e_{in}} $  &  Equivalent input-referred  voltage noise   \\
$e_n$  & Equivalent noise voltage    \\
$ e_{th}$  &  Thermal noise voltage generated by the source impedance   \\
$ E$ &   Young's modulus for the bacterial shell \\
$E_{\rm X}$  & Nernst potential  for ion X  \\
$ E(f)$ &  Electric field strength at  frequency  $f$  \\
$f$  & Frequency      \\
$f_o$  & Reference frequency of the lock-in amplifier    \\

$ F$ & Faraday's constant     \\
$I$ &  Applied ac bias current: $I=I_R + I_C $\\
$I_{in}$ &  Current flowing into the amplifier input  \\
$I_C $  &  Part of $I$ flowing through  $C_m$   \\
$I_R $  &  Part of $I$ flowing through  $R_m$  \\
$ I_1 $ &   Current passing through a single microchannel  \\
$ k$ &    Thermal conductivity of glass \\
$l$, $w$,  $h$ &   Length, width, height of a single microchannel \\
$l_b$ & Length of a bacterium \\
$ l_{cyl}$, $ r_{cyl}$, $ t$ &  Dimensions of a  bacterium in numerical models (a hollow cylinder with length $l_{cyl}$, radius $r_{cyl}$, and wall thickness $t$)  \\
$ l_\mu$ &  Cation drift length in a microchannel \\
$ \dot{m}$ &   Mass flow rate in a microchannel  \\
$ n$  &  Total number of charge carriers in the microfluidic resistor \\
$ { n_{\rm X}}$  &  Total number of intracellular $\rm X$ ions   \\
$N$  & Number of bacterial cells trapped in the microﬂuidic resistor    \\
$ N_1$ &  Number of bacteria   trapped in the  microchannel  \\
$ p_T$ &  Turgor pressure inside a bacterium  \\
$ p_{\rm {X}}$ &  Relative membrane permeability   \\
$\rm PSD$ & Power spectral density\\
$ q''$ &    Heat flux \\
$ Q$ &  Quality factor for the vibrational mode of a bacterium  \\
$ r(t) $ &  The rms value of resistance fluctuations (change)  \\
$ R$ & The universal gas constant     \\
$ R_{in}$  &   Equivalent resistance of the lock-in amplifier  \\
$R_m$  & Equivalent resistance of the microﬂuidic resistor    \\
$R_{n} $  &   Equivalent noise resistance representing all the white thermal noise sources\\
$ R_1$ &   Resistance of a single microchannel   \\
$ S(f)$  &   Normalized PSD of the excess voltage noise: $S(f) = {S_R(f) \over {R_m}^2}$\\
$ {S_{n}}(f)$  &     PSD of the fluctuations in the total number of charge carriers: ${{{S_R}(f)} \over {{{R_m}^2}}} = {{{S_n}(f)} \over {{n}^2}}$\\
$ {S_{n_{\rm X}}}(f) $  &    PSD for the  fluctuations of the number ${ n_{\rm X}}$ of  intracellular ions of type $\rm X$ in a single bacterium \\
$ S_R(f)$  &   PSD of the resistance fluctuations   \\
$ {S_R^{(1)}}(f)$ &    PSD of the resistance fluctuations  of a single microchannel  \\
$S_\rho $ &    PSD of the resistivity fluctuations  \\
$ S_V^{(ex)}(f,I)$  &   PSD of the current-dependent  excess voltage noise   \\
$ {\overline{S_V^{(ex)}(f)}}$  &   Averaged PSD  of the excess voltage noise: ${\overline{S_V^{(ex)}(f)}}  = {1\over {f_1-f_2}}{\int_{f_1}^{f_2} S_V^{(ex)}(f) df}$  \\
$ S_V^{(th)}(f,0)$  &  PSD of the (thermal) white noise      \\
$ S_V^{(tot)}(f)$  &    PSD of the  total voltage noise: $S_V^{(tot)}(f)=S_V^{(ex)}(f)+S_V^{(th)}(f,0)$  \\

%${\Re}$  &  The  real part of a complex impedance   \\
$ T$  &  Sample temperature    \\
$ T_{eff}$  &  Active  temperature of the bacteria  \\
$ T_{C}$ &  Microchannel temperature   \\
$ T_\infty$ &  The temperature far removed from the channel   \\
$ u_{\Delta p}$ & Pressure-driven component of the flow velocity in the microchannel \\
$ u_{EK}$ &   Electrokinetic component of the flow velocity in the microchannel\\
$ U$ &   Strain energy stored in a deformed bacterium  \\
%$ r_{300}$ &  The rms value of the resistance fluctuations caused by 300 bacteria  \\
$v(t)$  &  Time-domain voltage ﬂuctuations    \\
$ V_b$ &   Volume of a single bacterial cell \\
$V_{mem}$  & Electrical potential of the bacterial membrane   \\
$V_s$  & Reference oscillator output of the lock-in amplifier    \\
$V_A$, $V_B$ &  Voltages at the two diﬀerential inputs $A$ and $B$ of the lock-in amplifier \\
$ [{\rm X}]_i$ & Intracellular concentration of the ion  $\rm X$ \\
$ [{\rm X}]_o$ & Extracellular concentration of the ion $\rm X$    \\
$Z_c$  & Contact impedance at each of the four probes  in the microchannel device (microfluidic resistor)  \\
$Z_{in}$  & Equivalent impedance of the lock-in amplifier input circuit:  $Z_{in}= R_{in} \parallel C_{in}$   \\
$Z_m$  & Equivalent impedance of the microﬂuidic resistor: $Z_m= R_m\parallel C_m$    \\ \\

$\Gamma $  &  Coefficient of $I^2$ dependence of $S_V^{(ex)}(f,I)$:  $\overline{S_V^{(ex)}(f)}=\Gamma I^2$  \\
$\Delta f$  &  Measurement bandwidth   \\
$ {\Delta n_{rms}}$ &The rms value of the fluctuations in the number of charge carriers in the entire microfluidic resistor \\
$ {\Delta n^{(1)}_{rms}} $  &  $ {\Delta n_{rms}}$ per cell   \\
$\Delta p$  & Pressure drop along the microchannel during  bulk flow    \\
$ {\Delta [{\rm X}]_i}$ &   The rms fluctuations in the intracellular ion concentration of  ion $\rm X$  \\
$ \Delta \Psi$ &   Change in membrane potential due to the applied electric field  \\
$ \mu$ &   Mobility for small cations  \\
$ \nu$ &  Poisson's ratio for the bacterial shell  \\
$ \rho$ &  Electrical resistivity of the electrolyte \\
$ \rho_b$ &    Density for for the bacterial shell\\
$ {\phi}$  &  Rate of  transport of $\rm X$ ions through the cell membrane   \\
$ \tau$  & Overall effective relaxation time constant     \\
$ \tau_d$ &  Diffusion time constant for cations in a microchannel\\
$ \tau_f$ &   Flow time constant  in a microchannel  \\
$ {\tau _{\rm X}}$  & Lifetime for ions of $\rm X$ (or the relaxation time for a perturbation for by $\rm X$ ions within the cell)    \\
$ \omega_n$ &  Eigenfrequency  of a bacterium \\

\end{longtable*}

%\end{ruledtabular}
\end{center}

\bibliography{Noise.bib}% Produces the bibliography via BibTeX.

\end{document}

\title{Supplemental Material for\\ ``Measurement of the Low-Frequency Charge Noise of Bacteria''}

\newcommand{\BU}{Department of Mechanical Engineering, Division of Materials Science and Engineering, and the Photonics Center, Boston University, Boston, Massachusetts 02215, United States}

\author{Yichao Yang}
\affiliation{\BU}

\author{Hagen Gress}
\affiliation{\BU}

\author{Kamil L. Ekinci}
\email[Electronic mail: ]{ekinci@bu.edu}
\affiliation{\BU}

\date{\today}
\maketitle

\tableofcontents

\newpage

\section{Additional Data}

In this Supplemental Material, we provide additional data plots from control measurements. For the noise data shown in Sections A-E below, we calculate and correct the normalized PSD $S(f)$ for the excess noise using Eq. (5) in the main text. The noise data below for no bacteria, dead bacteria, and live bacteria are structured as follows. In Sections A and B, we show the noise measured in electrolytes  with no bacteria, including in LB and PBS at $37^{\circ}$C and LB at $23^{\circ}$C. In Section C, we  show the noise measured in electrolytes containing dead bacteria at $37^{\circ}$C. For each plot in Sections A-C, we show $S(f)$ for each current $I$ using small symbols, while we show the average $S(f)$  using large symbols. In Section D, we compare the noise from experiments with no bacteria with those  with dead bacteria. Our goal here is to establish that  the (background) noise level in just the electrolyte taken with no bacteria  is close to the noise level in the electrolyte containing dead bacteria.  In Section E, we show the noise measured with live bacteria at $37^{\circ}$C, with small symbols showing $S(f)$ for each $I$ and large symbols showing the average $S(f)$. In Section F, we present values of relevant parameters for different electrolytes in all measurements. In Section G, we show the normalized probability density functions of the voltage fluctuations in LB with no bacteria, dead bacteria, and live bacteria. All the relevant information is given in the captions. 

\subsection{Noise in Electrolytes Containing No Bacteria at $37^{\circ}$C}

  %*************************************Figure************************
\begin{figure*}[!h]
    \begin{center}
    \includegraphics[width=6.75in]{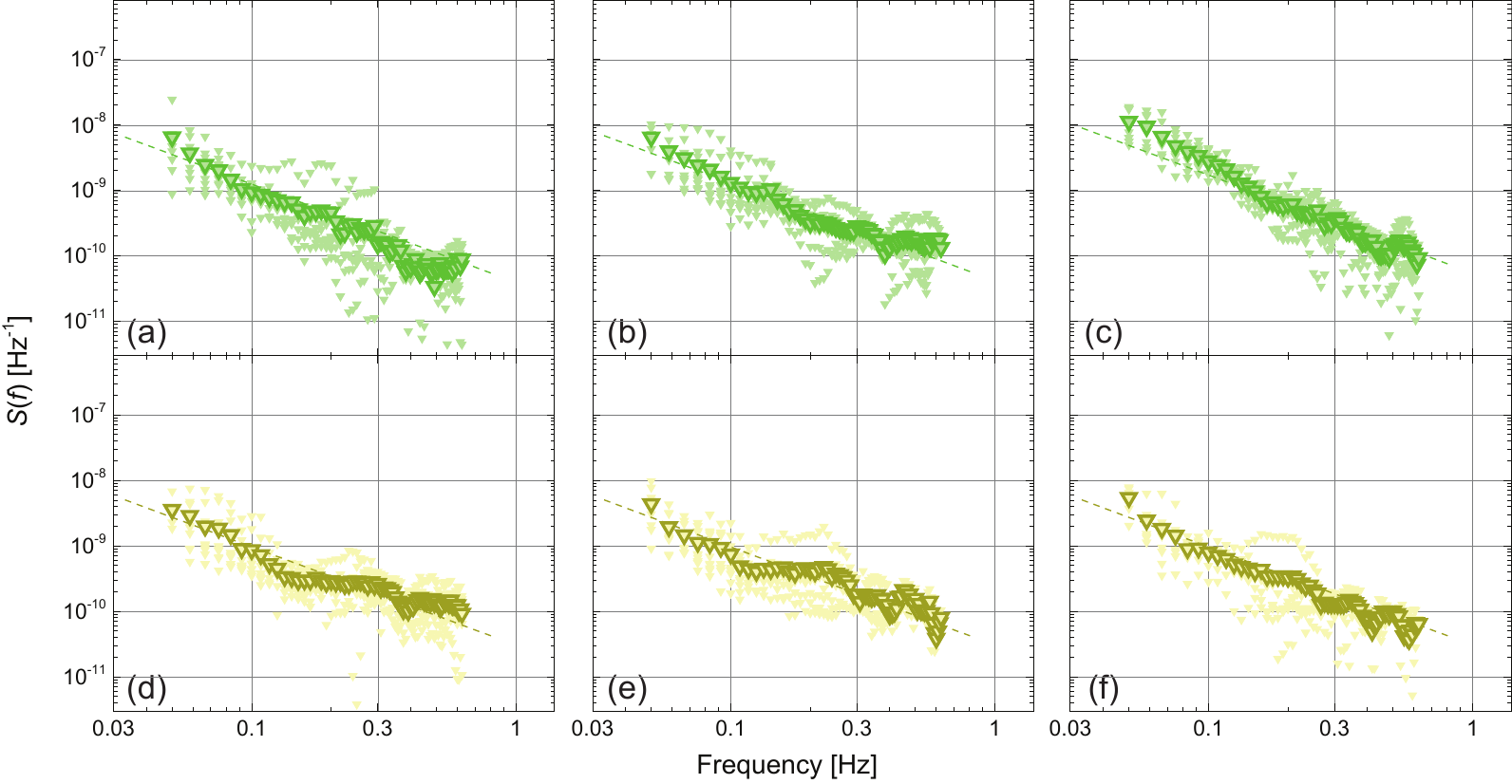} \\ %place holder
\caption{Normalized PSDs ${S(f)}$ of the resistance fluctuations for LB (a-c) and for PBS (d-f) containing no bacteria at $37^{\circ}$C. Each plot represents one independent experiment. In each plot, the small symbols show the data for each bias current, and the large symbols are the average of the data with different bias currents. The dashed lines correspond to $S(f)=B/f^{\beta}$. The values of $B$ and $\beta$ are given in Table \ref{tab_model_Parameters}.}
    \label{fig:figSPSDsno37}
    \end{center}
\end{figure*}
%%%%%%%%%%%%%%%%%%%%%%%%%%%%%%%%%%%%%%%%%%%%%%%%%%%%%%%%%%%%%%%%%%%

\clearpage

\subsection{Noise in Electrolytes Containing No Bacteria at Room Temperature}

  %*************************************Figure************************
\begin{figure*}[!h]
    \begin{center}
    \includegraphics[width=6.75in]{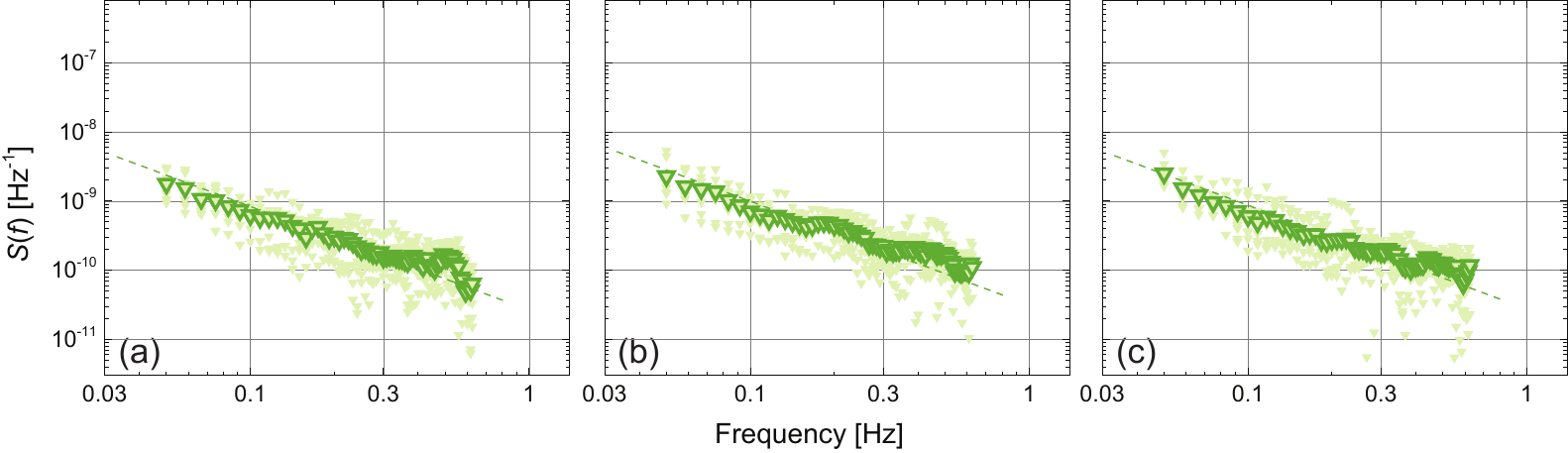} \\ %place holder
\caption{(a-c) Normalized PSDs ${S(f)}$ of the resistance fluctuations for LB containing no bacteria at $23^{\circ}$C. Each plot represents one independent experiment. In each plot, the small symbols show the data for each bias current, and the large symbols are the average of the data with different bias currents. The dashed lines correspond to $S(f)=B/f^{\beta}$. The values of $B$ and $\beta$ are given in Table \ref{tab_model_Parameters}.}
    \label{fig:figSPSDsno23}
    \end{center}
\end{figure*}
%%%%%%%%%%%%%%%%%%%%%%%%%%%%%%%%%%%%%%%%%%%%%%%%%%%%%%%%%%%%%%%%%%%

\subsection{Noise in Electrolytes Containing Dead Bacteria at $37^{\circ}$C}

  %*************************************Figure************************
\begin{figure*}[!h]
    \begin{center}
    \includegraphics[width=6.75in]{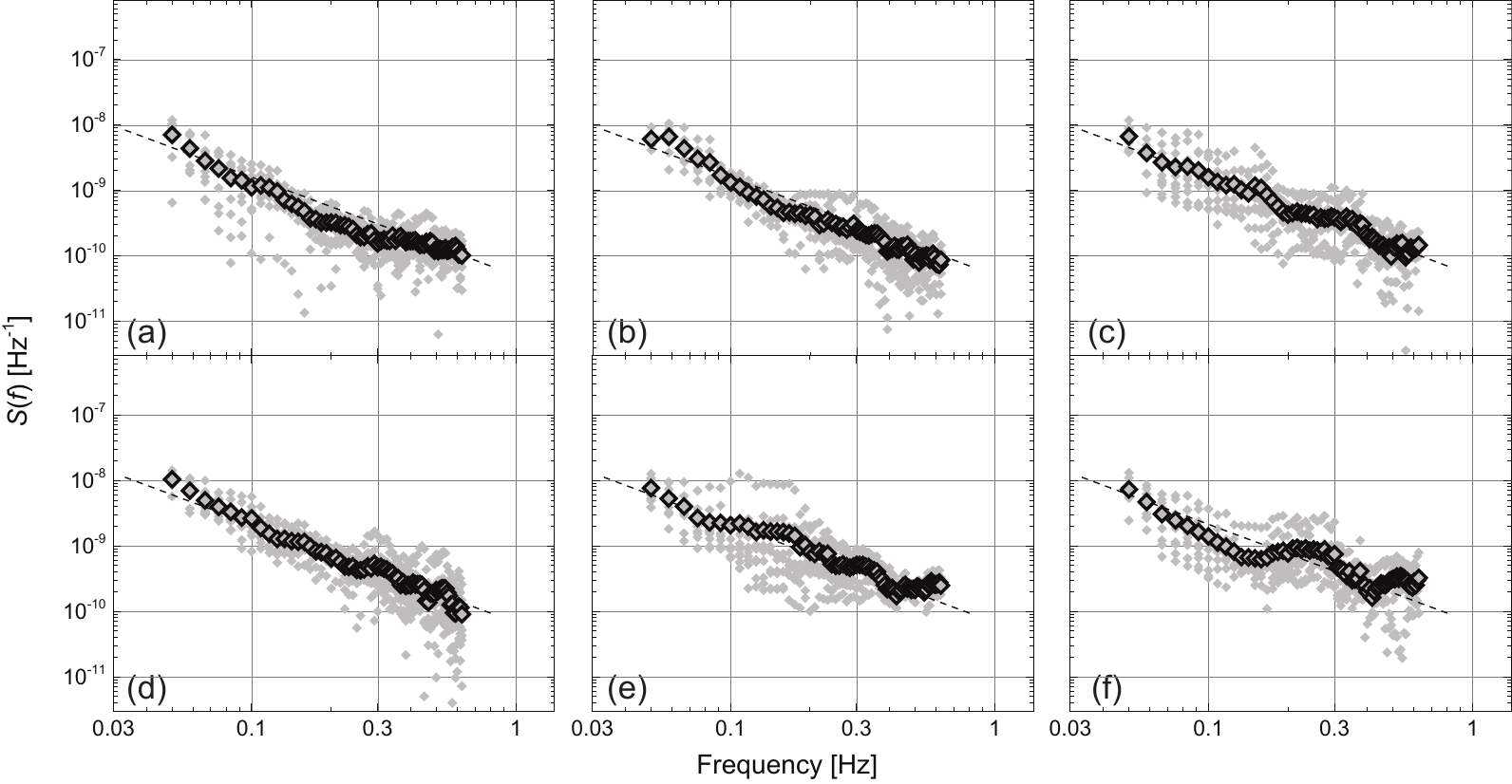} \\ %place holder
\caption{Normalized PSDs ${S(f)}$ of the resistance fluctuations for LB containing dead \textit{K. pneumoniae}  (a-c) and dead \textit{S. saprophyticus}  (d-f) at $37^{\circ}$C. Each plot represents one independent experiment. In each plot, the small symbols show the data for each bias current, and the large symbols are the average of the data with different bias currents. The dashed lines correspond to $S(f)=B/f^{\beta}$. The values of $B$ and $\beta$ are given in Table \ref{tab_model_Parameters}.}
    \label{fig:figSPSDsdead}
    \end{center}
\end{figure*}
%%%%%%%%%%%%%%%%%%%%%%%%%%%%%%%%%%%%%%%%%%%%%%%%%%%%%%%%%%%%%%%%%%%

\subsection{Summary of Background Data: Average Noise Curves in Different Electrolytes With and  Without Dead Bacteria}

Our experiments show different resistance values for microfluidic resistors filled with different electrolytes (containing no bacteria). At $37^{\circ}$C, the  $R_m$ values for LB and PBS are measured as approximately $2.5~\rm M \Omega$ and $1.7~\rm M \Omega$, respectively. At room temperature ($23^{\circ}$C), the measured $R_m$ value for LB  is approximately $3.0~\rm M \Omega$. The different $R_m$ values measured in PBS and LB are due to their different resistivities.  Based on the analysis  in the main text, we use  ${\cal C}(R_m)$ to correct the measured noise when making a comparison of the excess noise levels between different electrolytes. 

\begin{figure}[!h]
    \begin{center}
    \includegraphics[width=3.375in]{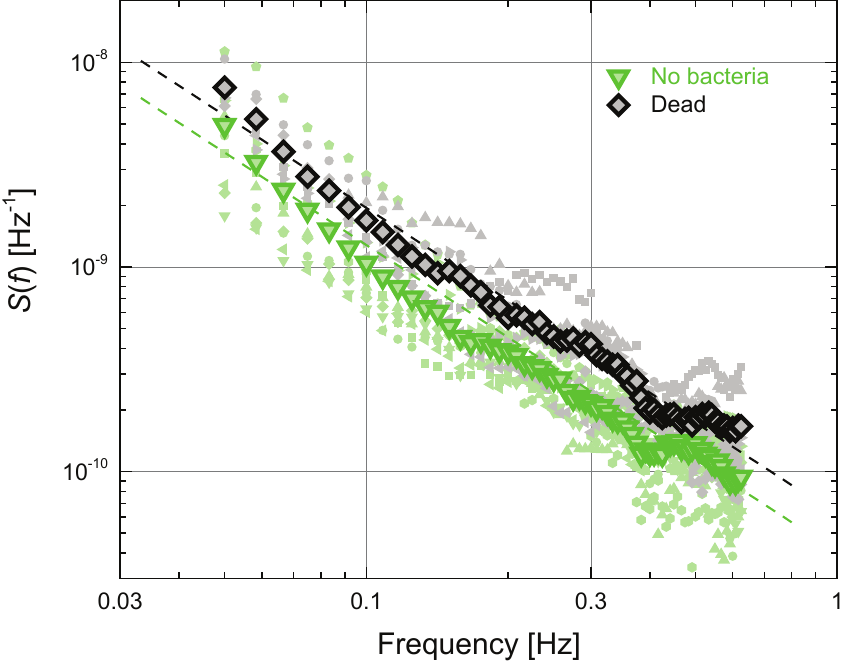}  %place holder
    \caption{Normalized PSD $S(f)$ of the resistance fluctuations for electrolytes containing no bacteria and dead bacteria. The small green  symbols show data from nine independent experiments, three each for PBS at $37^{\circ}$C, LB at $37^{\circ}$C, and  LB at $23^{\circ}$C. The small gray symbols show data from six independent experiments three with dead \textit{K. pneumoniae} and three with dead \textit{S. saprophyticus}, both in LB at $37^{\circ}$C. The larger symbols show the average values of the experiments with no bacteria (green) and with dead bacteria (black). The dashed lines are not fits but show the asymptotic behavior and correspond to ${S(f)}={B/f^{\beta}}$, with $B = 4.05 \times {10}^{-11}$ and $\beta = 1.50$ for no bacteria (green) and  $B = 6.15 \times {10}^{-11}$ and $\beta = 1.50$ for dead bacteria (black).}
    \label{fig:nodead}
    \end{center}
\end{figure}

Figure \ref{fig:nodead} shows a comparison between the normalized PSD, $S(f)$, of the noise measured from electrolytes containing no bacteria (green) and dead bacteria (black). Here, the data for no bacteria (at $37^{\circ}$C and $23^{\circ}$C) are the average of $S(f)$ for each $I$ value shown in each plot of Figs. \ref{fig:figSPSDsno37} and  \ref{fig:figSPSDsno23}; the data for dead bacteria (at $37^{\circ}$C) are the average of $S(f)$ for each $I$ shown in each plot of Fig. \ref{fig:figSPSDsdead}. We calculate and correct the curve for $S(f)$ in  Figs. \ref{fig:figSPSDsno37}-\ref{fig:figSPSDsdead} for each $I$ using Eq. (5) in the main text; the value of ${\cal C}(R_m)$ for each $R_m$ is plotted in Fig. 9 in the main text; the $R_m$ value for each experiment is listed in Table (\ref{tab:tabS2}). We find that the $S(f)$ curves in Fig. \ref{fig:nodead} for no bacteria and dead bacteria exhibit very close noise levels and follow the same scaling behavior of the form ${S(f)}={B/f^{\beta}}$.

 %*************************************Figure************************
%%%%%%%%%%%%%%%%%%%%%%%%%%%%%%%%%%%%%%%%%%%%%%%%%%%%%%%%%%%%%%%%%%%

\clearpage
\subsection{Noise in Electrolytes Containing Live Bacteria at $37^{\circ}$C}

  %*************************************Figure************************
\begin{figure*}[!h]
    \begin{center}
    \includegraphics[width=6.75in]{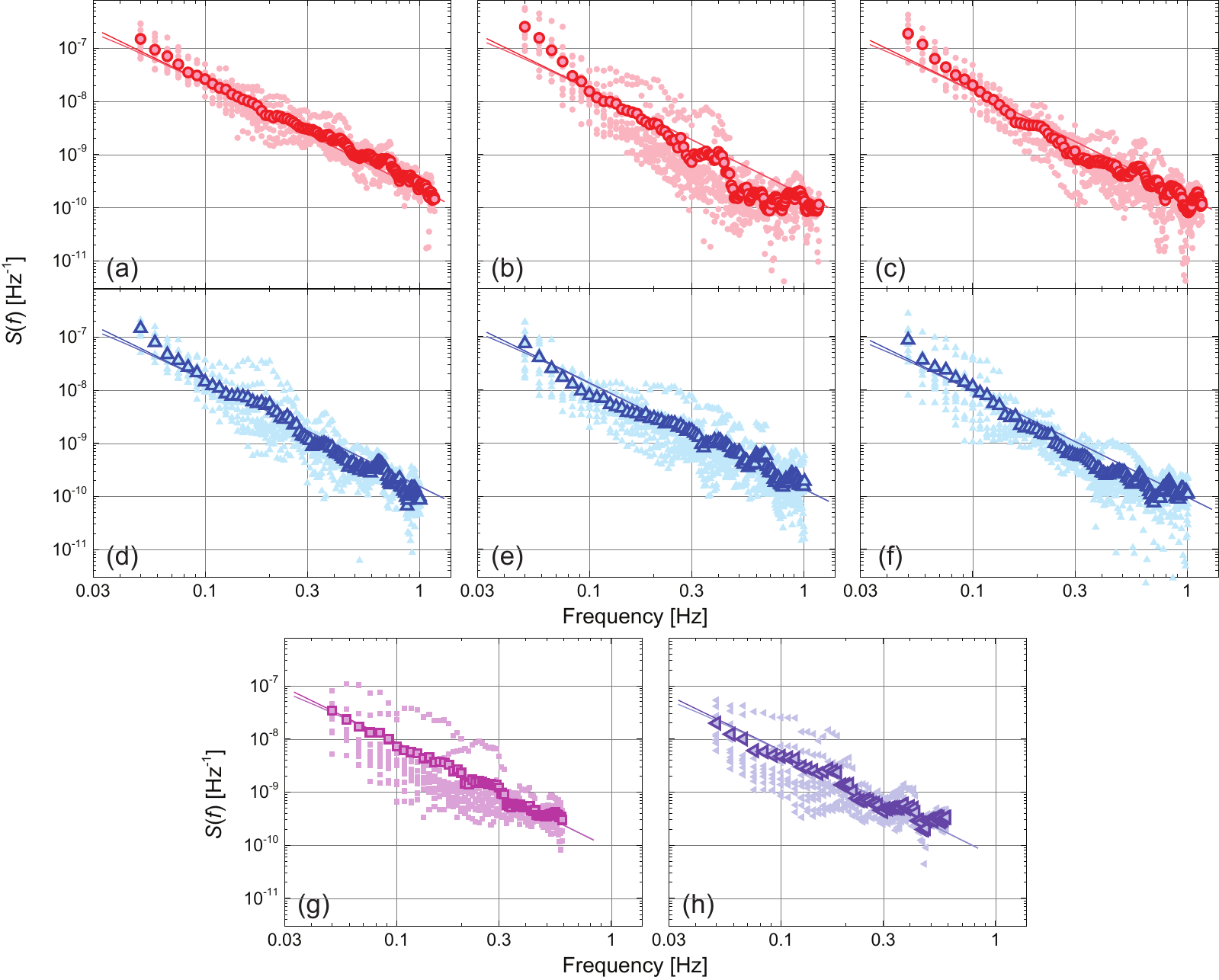} \\ %place holder
\caption{Normalized PSDs ${S(f)}$ of the resistance fluctuations of LB containing  live \textit{K. pneumoniae}  (a-c) and live \textit{S. saprophyticus}  (d-f), and of PBS with live \textit{K. pneumoniae} (g) and live \textit{S. saprophyticus}  (h) at $37^{\circ}$C. Each plot represents one independent experiment. In each plot, the small symbols show the data for each bias current, and the large symbols are the average of the data with different bias currents; the solid  curves   correspond to $S(f)= {{A \tau ^2 } \over {1 + 4{\pi ^2}{f^2}{{\tau} ^2}}}$, with $\tau=10~\rm s$ (bottom), and with $\tau=30~\rm s$ (top); the values of $A$ corresponding to  different $\tau$ values are given in Table \ref{tab_model_Parameters}.}
    \label{fig:figSPSDslive}
    \end{center}
\end{figure*}
%%%%%%%%%%%%%%%%%%%%%%%%%%%%%%%%%%%%%%%%%%%%%%%%%%%%%%%%%%%%%%%%%%%

\clearpage
\subsection{ Values of Relevant Parameters  for Different Electrolytes}

%*************************************TABLE***************************************
\begin{table}[!h]
\caption{\label{tab_model_Parameters} Summary of the values of the parameters  in all measurements. The first and second columns show the electrolytes and temperatures used in the experiments, respectively. The third column lists the measured bacteria: \textit{K. pneumoniae} (KP) and  \textit{S. saprophyticus} (SS). The fourth column shows the values of $R_m$ with $N\approx 300\pm50$ of (dead or live) bacteria in the microchannel. The fifth column shows the values of the electrical conductivity for the electrolytes. The sixth and seventh columns show the values of $A$ and $\tau$ for live bacteria in the formula $S(f) = {{A \tau ^2} \over {1 + 4{\pi ^2}{f^2}{{\tau} ^2}}}$. The last two columns show the values of $B$ and $\beta$ for no bacteria and dead bacteria in the formula $S(f) = B/f^{\beta}$.}
\begin{ruledtabular}
\begin{tabular}{cccccccccc}

Electrolytes & Temperature & Bacteria & Typical $R_m$ & Conductivity & $A$  & $\tau$ & $B$ & $\beta$ \\
 & [$^{\circ}$C] &  & [$\rm M \Omega$] & [S/m] & [$\times {10}^{-9}$]  & [s] & [$\times {10}^{-11}$] &  \\
     
\hline

$\rm LB$  & 37       & -    &  $2.50\pm0.04$ &1.00&  -  & - & 3.95, 4.15, 5.45    & 1.50, 1.50, 1.50   \\
$\rm LB$  & 23       & -    &  $3.00\pm0.04$ &0.83 &  - & -  & 2.65, 3.15, 2.75    & 1.50, 1.50, 1.50   \\
$\rm PBS$ & 37       & -    &  $1.70\pm0.002$ &1.47 &  - & -  & 3.05, 3.05, 3.05    & 1.50, 1.50, 1.50   \\
$\rm LB$ & 37 &  dead KP    &  $3.10\pm0.40$ &-  &  - & - & 5.05, 5.05, 5.05   & 1.50, 1.50, 1.50   \\
$\rm LB$ & 37 &  dead SS &  $3.10\pm0.40$ &-  & - & - & 6.75, 6.75, 6.75 & 1.50, 1.50, 1.50  \\
$\rm LB$  & 37    &  live KP     & $3.10\pm0.40$&- & 8.80, 6.80, 6.30    & 10, 30   & - & -  \\
$\rm LB$  & 37    &  live SS  & $3.10\pm0.40$ & & 6.10, 5.40,  3.80  & 10, 30   & - & - \\
$\rm PBS$ & 37    &  live KP     &  2.15&- &  3.40&  10, 30   & - & -   \\
$\rm PBS$  & 37   &  live SS  & 2.24&- &   2.40&  10, 30   & - & -  \\

\label{tab:tabS2}
\end{tabular}
\end{ruledtabular}
\end{table}
%****************************************************************************
%****************************************************************************

\subsection{Probability Density Functions of Voltage Fluctuations}

  %*************************************Figure************************
\begin{figure*}[!h]
    \begin{center}
    \includegraphics[width=6.75in]{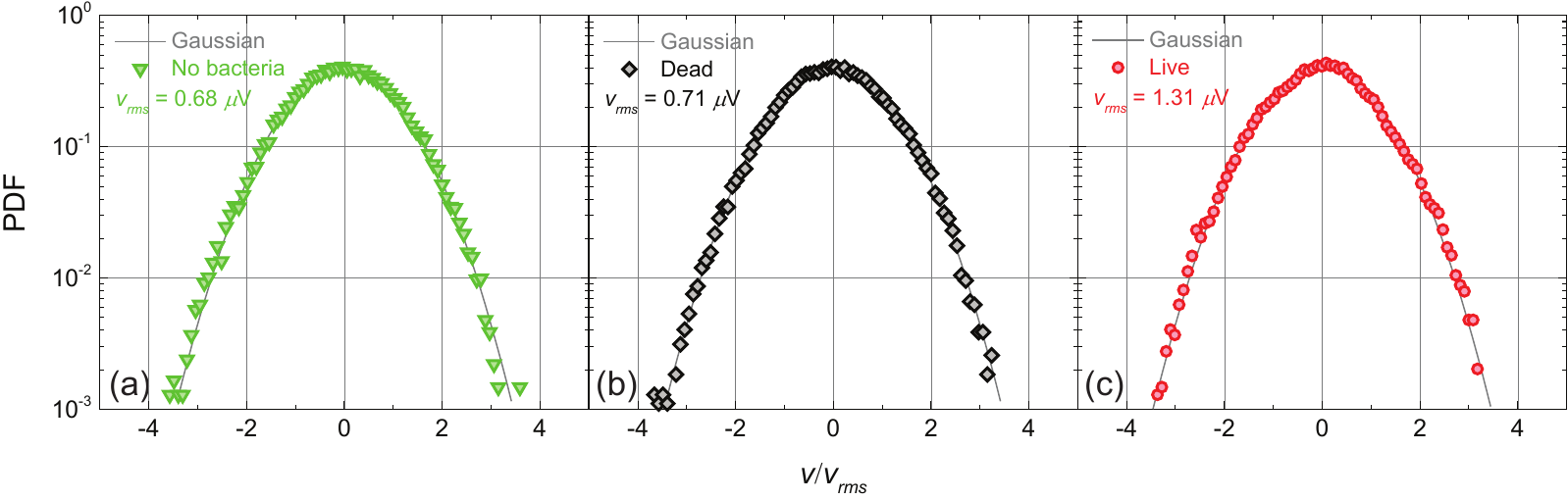} \\ %place holder
\caption{Normalized probability density functions (PDFs) for the voltage fluctuations plotted in units of the rms values $v_{rms}$  in LB containing no bacteria (a),  dead bacteria (b), and live bacteria (c) at $37^{\circ}$C. Here, the  voltage fluctuations are all measured using $I\approx 7.80~\rm nA$; the data of live and dead bacteria are  \textit{K. pneumoniae}; the gray line in each plot is a Gaussian; the value of $v_{rms}$ for each data set is noted in the figure.  For computing the PDFs shown in (a-c),  we use four time series of  $v(t)$ (each of 120 s) for the same experiment, such as those shown in Fig. 3(b) in the main text. We  calculate the values of the normalized $n$-th moments of the PDFs, ${\mu_n} \over {\sigma^n}$, where ${\mu_n}=\langle v^n \rangle$ and ${\sigma}=\langle v^2 \rangle^{1/2}$ (the  brackets indicate averages over the PDF) are respectively the $n$-th central moment and the standard deviation of the voltage fluctuations. We find the normalized high-order moments as 3.02 ($n=4$), 15.01 ($n=6$), and 102.17 ($n=8$) for no bacteria (a), 2.98 ($n=4$), 14.48 ($n=6$), and 95.61 ($n=8$) for dead bacteria (b), and 3.05 ($n=4$), 14.76 ($n=6$), and 93.00 ($n=8$) for live bacteria (c).}
    \label{fig:figPDFs}
    \end{center}
\end{figure*}
%%%%%%%%%%%%%%%%%%%%%%%%%%%%%%%%%%%%%%%%%%%%%%%%%%%%%%%%%%%%%%%%%%%

%\clearpage
%\bibliography{Noise.bib}